\documentclass[12pt]{article}
\usepackage{graphicx}
\usepackage[square,numbers]{natbib}
\usepackage[colorlinks=true, linkcolor=blue, citecolor=blue, urlcolor=blue]{hyperref}

\usepackage{amsmath}
\usepackage{authblk}
\usepackage{setspace}
\usepackage[top=1.5in, bottom=1.5in, left=1in, right=1in]{geometry}

\begin{document}
\title{Identification of phase correlations in Financial Stock Market Turbulence}
\author[1]{Kiran Sharma}\affil[1]{Department of Commerce, Sikkim University}
\author[2]{Abhijit Dutta}\affil[2]{Department of Commerce, Sikkim University}
\author[3]{Rupak Mukherjee}\affil[3]{Department of Physics, Sikkim University}

\date{\today}

\maketitle
\begin{abstract}
    The basis of arbitrage methods depends on the circulation of information within the framework of the financial market. Following the work of Modigliani and Miller, it has become a vital part of discussions related to the study of financial networks and predictions. The emergence of the efficient market hypothesis by Fama, Fisher, Jensen and Roll in the early 1970s opened up the door for discussion of information affecting the price in the market and thereby creating asymmetries and price distortion. Whenever the micro and macroeconomic factors change, there is a high probability of information asymmetry in the market, and this asymmetry of information creates turbulence in the market. The analysis and interpretation of turbulence caused by the differences in information is crucial in understanding the nature of the stock market using price patterns and fluctuations. Even so, the traditional approaches are not capable of analyzing the cyclical price fluctuations outside the realm of wave structures of securities prices, and a proper and effective technique to assess the nature of the Financial market. Consequently, the analysis of the price fluctuations by applying the theories and computational techniques of mathematical physics ensures that such cycles are disintegrated, and the outcome of decomposed cycles is elucidated to understand the impression of the information on the genesis and discovery of price and to assess the nature of stock market turbulence. In this regard, the paper will provide a framework of Spectrum analysis that decomposes the pricing patterns and is capable of determining the pricing behavior, eventually assisting in examining the nature of turbulence in the National Stock Exchange of India. 
\end{abstract}

\newpage

\section{Introduction}
The Financial market is extremely reactive to changes in the market environment. A small information on underlying factors will lead to changes in market behavior and the pricing patterns. The information which are directly or indirectly related to the market affects the price movement and is vital for price discovery. Reflection of companies' performance and overall dynamics of the market on price provides a foundation for the stakeholders to assess trends and try to discover the price to make informed decisions. The traces of price discovery go back to the 1900s when Bachelier talked about the application of probability theory to analyze the price and gain from options through his work “Theory of speculation”. Followed by Fama (1965)(1970)\citep{fama1965}\citep{fama1970} laid the foundation for the efficient market hypothesis, arguing that prices are random and follow a random walk. He also identified three forms of market efficiency where past prices, public information, and private information play a role in determining the prices. But it is not possible to access all the available information, and no one is capable of accessing all the information, resulting in a random nature in price fluctuations and patterns. The study by Fama et al (1969)\citep{fama1969} revealed that the market prepares for splits, adjusts the prices accordingly, and is capable of fully incorporating the information, thus supporting the Efficient Market Hypothesis. Laffont \& Maskin(1990) \citep{laffont1990} investigates whether the Efficient Market Hypothesis (EMH) holds under two conditions: first, where prices do not reveal private information, and second, where prices do. They analyze how large traders mask the information using Perfect Bayesian Equilibrium and comparative statistics. They found that the market reveals the information, but at the same time, the traders do not prefer to have the market as efficient. This indicates that the EHM is not applicable in imperfect competition when a large group of traders hides information, preventing the price from reflecting the private information that they hold. Mitchel \& Mulherin (1994)\citep{mitchell1994} also found that the news reported in Dow Jones \& Company newspaper is positively correlated to the trading volumes and market returns of US securities, indicating a direct relation to the trading activity. Chan et al.(2008)\citep{chan2008} studied information asymmetry between foreign and domestic shares in the Chinese stock market. Their analyses suggested the presence of information asymmetry between foreign and domestic shares in the Chinese stock market. The study by Urquhart\&McGroarty (2016)\citep{urquhart2016} found the traces of an efficient market on S\&P 500 and EUROSTOXX 50 by applying AR-GARCH and Variance. Due to the impact of information on price discovery and fluctuations., Fama’s theory of Efficient Market Hypothesis is still relevant and is important to analyze the stages of market patterns and fluctuations to discover the quantitative aspect of the information in terms of price dynamics.

The prices are impacted by all those direct or indirect factors, signaling that price fluctuations emerge from the changes and information in the market. Fama urged that the market incorporates all information in an efficient manner and the market are always efficient\citep{fama1970}. He also stated that the prices are always random and are capable of reflecting all available information,  and the pricing patterns are always unpredictable. Due to the importance of the role of information on price discovery and pricing patterns, it makes investors difficult task to predict the market with certainty. But the development of different theories like Arbitrage Pricing Theory(APT) and Capital Assets Pricing Models (CAPM) helps to understand how various factors influence the asset pricing, providing an advantage to the arbitrage \citep{ross1973}\citep{famaFinch2004}. Considering the information and the underlying factors, investors try to assess such trends and movements of price to make informed decisions, as such price reflects the overall performance of the listed companies and overall market dynamics. The rising stock prices indicate enhanced company performance, while declining prices indicate downward trends. Stock prices display cyclical movements that are influenced by the dynamic nature of micro as well as macroeconomic factors. followed by the reactions of investors to any economic factors can lead to variations in share prices\citep{martynova2008}. Each stock price exhibits unique cyclical patterns, which are studied separately or with the market. It is also important to observe that the cyclical patterns and behaviors of prices may have emerged from trends and autocorrelation\citep{aydogan1988}. To understand the market dynamics, one needs to analyze the pricing patterns and their cyclical fluctuations, which ultimately help to predict future price movements. Examining the cyclical components of time series data is important for understanding the market trends and forecasting the prices, supporting the interpretation of the movements, whether short or long period\citep{idrees2019}. Forecasting the market depends on identifying the patterns in price sequences by deeply analyzing the historical prices to improve predictability\citep{cowles1944stock}.

while discussing the price discovery. It is also equally important to recognize and understand that the patterns are not always uniform and are always influenced by the uncertainties and asymmetric information. Such interaction of the elements has offered the interest of examining the price assessment and prediction, ultimately leading an informed decision-making. A study by Malkiel (1989) \citep{malkiel1989} suggested that to have more research on the dynamics of pricing to discover the true nature of pricing patterns, whether they are truly because of the nature of the market or a result of data exploitation.
The fluctuations in the stock market are not always consistent and are impacted by various related factors, as well as events and information asymmetry, which causes turmoil in the market, signaling an unpredictable nature of the price movements. These regular ups and downs of the stock prices and continuous fluctuations of the movement of price are known as turbulence \citep{wang2020} \citep{mantegna1999}. Such situations in the market create volatility, making it difficult to understand the cyclical patterns and movements of prices. The turbulence in the financial market is generally caused by the situation of information asymmetry and high-energy events, and it is important to analyze the market turbulence and related factors. Engel et al.(2018) \citep{engel2018} studied the major financial crisis, which caused the turbulence in the stock market and framed an early Warning signal using Markov switching models on such situations for Asian and European stock markets. Another study by Khoojine \& Han (2020)\citep{khoojine2020}  developed the SPNAR stock prices network auto-regressive model to analyze the Chinese stock market during pre- and post-turbulence. Understanding such situations of uncertainty is essential for market analysis and helps to prevent such situations. Due to the flow of asymmetric information and the dynamic nature of the market, the market becomes volatile with rapid fluctuations, which makes it difficult to analyze and predict the market through the price pattern. So the examination of cyclical pattern ( rapid ups and downs) of the market is valuable to understand the nature of the market as turbulent or not.  In this paper, we will analyze the Indian stock market, especially the National Stock Exchange of India, considering the factor of efficient market and the price dynamics through which the nature and state of turbulence within the NSE can be understood. The concepts and techniques of computational physics, especially the spectrum analysis, will be employed to understand the above-mentioned scenario, contributing a different approach for analyzing the market dynamics.
\subsection{Stock Market and Turbulence}
The stock market is impacted by unpredictable events and the turbulence caused by such events needs to be understood to predict and understand the impact of such events on the stock market. The stock market is unpredictable and due to the rapid changes in environment and the events which affects the financial market leads to disturbances and situation of volatility and such unpredictable ups and downs of the stock market is termed to be known as the state of Turbulence. Turbulence in the stock market refers to rapid and significant increase and decrease in the prices, followed by the situation of unpredictability in the financial market. Turbulence signifies the span of extreme volatility and instability in the financial market which are characterized by the fluctuations and huge level of uncertainty \citep{wang2020}. The word turbulent is commonly used as the fluctuations in price and patterns which replicate the velocity fluctuations in the turbulence \citep{mantegna1999}. 

Turbulence can be viewed as a combination of various independent cycles/frequencies present in a system. Typically, in fluid turbulence, a fluid medium can sustain several independent frequencies at the same time. This happens as fluids cannot sustain strong shear forces and fluid elements try to thermalise through conduction as well as convection mechanism. When energy is pumped into a fluid medium, typically the energy resides in large wavenumbers or in long-time cycles of the fluid’s motion. However, this energy cascades into shorter wavenumbers or in short-time cycles soon. This cascading of energy into shorter scales is called the inertial range of turbulence. Finally, the energy is drained out in the form of heat, through tiny viscous scales of the system. Thus, typically a fully-developed turbulent fluid contains three clear scales – the first being the scale at which the energy is being injected into the system, the second is the inertial scale where the energy cascades to the shorter scales and the third is the viscous scale where the energy gets drained out from the medium in the form of heat/radiation. The second phase, that is, the inertial scale, plays the key role in developing the turbulence in a fluid. This inertial scale transfers the energy (/amplitude in the stock market) from large-scale vertices (/long-time cycles) to smaller vertices (/short-time cycles or fluctuations) through some nonlinear interactions. It has been found that in a fully developed turbulence, these nonlinear interactions between different modes (/cycles) actually retain information from both the amplitude as well as the phase of each of the modes (/cycles). Thus, merely an analysis using the energy spectrum falls short in understanding the actual nature of the turbulence present in the system. For example, inside a fully developed turbulence, we expect there will be no correlation between the individual phases, and thus each cycle will have its own unique phase information. In addition, if there is a huge correlation between the phases of individual cycles, one can conclude that the medium has not reached a fully developed turbulent phase. In this report, we describe a method to dig out this phase information from stock market data (using an integral transform method) and check for the quality or degree of turbulence in these individual data. This method is already in use to identify and quantify the level of turbulence in fluids as well as in plasmas \citep{kumar2019} \citep{kumar2016FFFF}. However, to the best of our knowledge, this is the first time we employ this technique to categorize turbulence in stock market data. \citep{nussbaumer1982fast}\citep{kumar2019observation}
\subsection{Market dynamics and Computational Approach}
Due to the effectiveness of fundamental analysis to examine the stock market for long-term investors, it lacks effectiveness on short-term predictions as it considers a range of variables, including macroeconomic factors, industry dynamics, and company-specific elements, to determine the value of financial assets. However, this analysis can be complex, requiring consideration of multiple variables and significant time investment, making it more suitable for long-term investors. \citep{graham1951}, \citep{graham2005intelligent}. In contrast, due to the desire for quicker gains and faster analysis of market trends, technical analysis emerged. This approach posits that all available information is already reflected in stock prices, allowing analysts to focus on price trends using historical data to capitalize on market movements \citep{hamilton2006}, \citep{dow1}, \citep{dow2}, \citep{Rhea}, \citep{griffioen2003}.
Technical analysis has experienced significant growth, particularly among short-term traders. Due to the complexity, noisy, chaotic, volatile, and vast amount of data that needs to be analyzed, machine-learning techniques have emerged and delivered superior results. The computational approach, generally a technical analysis, provides more meticulous outcomes and the ability to analyze big data. The data analysts nowadays are using computational and machine learning to analyze financial data. The stock market data are very complicated and non-stationary, making the traditional approaches typically inaccurate, and they treat stock data as a linear time series. The study by Kumar et al.(2021)\citep{kumar2021} considered such issues and reviewed the recent developments on the application of computational methods in the financial markets. Their study found that methods like Artificial Neural Networks, Support Vector Machines are intensively used in place of the traditional approach of data analysis. Similarly, Dunis et al. (2014) \citep{dunis2014} also provided different computational techniques that are used for analyzing the stock data. Sonkavde et al. (2023)\citep{sonkavde2023} presented a comprehensive analysis of various algorithms, such as time series and deep learning techniques, for stock market predictions and analysis. They discussed the use of methods like linear regression, ARIMA, and other approaches under the computational framework with a robust result. In the context of technical analysis or computational approach, there exist different indicators, like chart patterns, phases, and candlesticks, which are accepted as fundamentally significant and efficient. Li\&Bastos(2020)\citep{li2020} and Fauzi\&Wahyudi(2016) \citep{fauzi2016}stated that image processing and regression techniques will further enhance these analyses. Their findings indicate that the technical analysis can yield more accurate results for market forecasting and price prediction. Taking into account of effectiveness of the computational approach, this study highlights the importance of applying the computational approach to examine the financial data.  This study will apply the integral Fourier Transforms, which are especially employed in the fields of physics and engineering. The Fourier transform is used to break down the time series data, generally on signals, to convert them from the time domain to the frequency domain. This mathematical technique is applied to analyze and represent the signals by breaking them into frequency elements.

Fourier Transform (FT) is a mathematical method for analyzing time series data and decomposing it into frequency components, especially used to analyze the cyclical movements. FT permits spectral analysis, providing a band of time series data and helping to disclose the dominant frequencies and patterns, like seasonal trends or business cycles. The properties of the Fourier Transform allow stacking the signals into the frequency domain, which simplifies the analysis to understand their interactions to form the signal. This mathematical approach has multiple applications comprising signal processing, engineering, and economics, which enhances our understanding of the cyclical patterns and phases \citep{cochran1967}. The mathematical approach of the Fourier Transform and considering its effectiveness in analyzing the cyclical patterns of the time series data will be employed to explore the National Stock Exchange Data.

Here in this project,the  we use FFTW library \citep{FFTW05}(@https://www.fftw.org/) within a computational architecture named TARA(@https://rupakmukherjee.github.io/TARA/) \citep{mukherjee2019.41}\citep{mukherjee2019.42}\citep{mukherjee2018.43}\citep{mukherjee2018.46}. The TARA simulation framework is a multi-dimensional pseudo-spectral solver to simulate incompressible and weakly compressible fluid or plasma turbulence \citep{mukherjee2019.44}. TARA has the flexibility to add higher-order fluid-moment equations with minimal computational overhead \citep{mukherjee2019.48} \citep{mukherjee2018.50}. This architecture runs efficiently on massively parallel CPU as well as multi-GPU architectures \citep{mukherjee2019.49} \citep{mukherjee2019.43}. In addition, the performance scales efficiently under MPI as well as OpenMP (shared- and distributed-memory) clusters \citep{biswas2021.51} \citep{biswas2020.52} \citep{gupta2019.47}. Here in this work, we extend the TARA diagnostic module and perform an extended form of extended integral transform analysis (Kim \& Powers, 1979) of time-series data. We benchmark our new module with existing analytical results and then perform an analysis of NSE data with our new module. 
\section{Research Gap}
Despite being effective in analyzing the tick data, the application of computational approaches is infrequent. additionally, the application of data analysis techniques from statistical/ mathematical physics is shockingly limited. The importance of using such methods yields superior results. The mathematical approach of analyzing the data through converting it to the frequency domain can enhance our understanding of the nature of frequencies, providing better and deeper insights into market dynamics related to efficiency and turbulence. Application of such techniques of analyzing the dynamics of the financial market through mathematical approaches, especially using integral transforms by converting the data from the time domain to the frequency domain, has not been explored in previous studies.
\section{Research objective}
Considering the dynamics of information and efficient markets, this proposed study will contribute to the financial market analysis through a computational approach. The primary objective of this paper is to lay out a mathematical approach to analyze the stock market using techniques of computational physics. Through integral analysis and visualization, this study assesses how asymmetric information affects the market behavior. Further, the phase relations and cyclical pattern of stock prices will be reviewed by using the concepts and methods of mathematical physics to magnify the understanding of phases and modes of cyclical patterns of stock prices, especially quantifying the degree of financial market turbulence. The framework of spectrum analysis, which this study introduces, is capable of understanding the nature of market turbulence in a deeper and critical way.
\section{Proposition}
\begin{enumerate}
    \item The behavior of price index of stock market can be decomposed into a periodic nature, with fluctuations on top of it. Thus a pattern analysis technique can lead to the prediction of market price rallies and become crucial for understanding the nature of the market. \citep{peters1996}\citep{sharma2002} \citep{akar2011} \citep{cootner1962} \citep{healy2001}
    \item When there is no coherent temporal pattern in the periodic nature, it is possible to segregate few independent periodic pattern of different timescale. However, this pattern can be used to surrogate specific information inside the repetitive pattern itself to control the predictibility of the market.
    \item Since standard integral transformations (for example Fourier transformation) cannot distinguish this information asymmetry, surrogated inside the periodic patterns generated from genuinely independent frequencies, there is a chance that the biased frequencies have been planted artificially to employ bias in the market. 
    \item In this paper, we employ an extended form of an integral (Fourier) transformation to distinguish between surrogated ‘phase correlations’ inside a periodic temporal fluctuation and uncorrelated periodic temporal fluctuations. 
    \item Further, we delineate our model for one dimensional direct numerical simulation data for Burgers turbulence and diffusion equation. Finally we use it to analyse the nature of the raw market data and its turbulence.
\end{enumerate}
\section{Research Methods}
This study provides an unconventional method to analyze phase relations between different frequencies of time series data. The criticality of this method is outlined in the Proposition. First, we apply our model to illustrative data, and then we apply it to real stock market data.
Given a function in the time domain, it is possible (given that it satisfies the Dirichlet criteria) to convert it to the frequency domain by taking a Fourier transform of the given function.
\begin{equation}
    F(\omega) = \frac{1}{\sqrt{2\pi}} \int_{-\infty}^{\infty} f(t) \, e^{i\omega t} \, dt
\end{equation}

Here $f(t)$ denotes the function in time domain and the function $F(\omega)$ denotes the corresponding Fourier transform in frequency $(\omega)$ domain. We can extend this integral transform into cases of discrete datasets as well in the following way:
\begin{equation}
    F(\omega) = \sum_{t=0}^{N-1} f(t) \cdot e^{-i \cdot \frac{2\pi \omega t}{N}}
\end{equation}

The amplitudes of the Fourier modes denote the strength of the corresponding frequency present in the timeseries data $(f(t))$. Each one of the frequencies represent one cyclical pattern movement of the data. When there are multiple frequencies present, the cyclical movement becomes cumbersome and requires an in-depth analysis. In our analysis we use a numerical library: FFTW (Fastest Fourier Transform in the West) to evaluate the Fourier transform of both illustrative data as well as the stock market data.
However, the discrete Fourier transform of data (in real numbers) generate Fourier modes in complex number domain. It is usual practice to evaluate the square norm of the complex data to draw inferences, as a square norm maps the data back to the real-number space. However, this method throws away the phase relations between different complex Fourier modes and hence we may potentially loose the valuable information about the phases and frequencies between the modes which are hidden inside the data. 
In this analysis, we propose an extension of this integral transform (in this report, Fourier transform) such that the phase relations are retained. Thus, an artificially planted information/frequency will appear as a hotspot in our extended integral transform, as it will maintain a fixed phase relationship with the naturally occurring cluster of independent frequencies. This extended form of Fourier Transform will be used to understand the phase relations and independent Fourier modes of stock prices, which will further help in understanding the market turbulence.

\subsection{Tools and Techniques}
To begin with, we'll create an illustrative sinusoidal dataset that has a fixed frequency and a fixed phase. This study will then introduce another sinusoidal dataset with a different fixed frequency and phase. The sinusoidal datasets are generated using a sine wave, which is a repeating wave pattern often described as oscillation. We create two copies of this same dataset. In one of the copies, further add another sinusoidal dataset of a superposition of the two frequencies and a third fixed phase. This third phase is independent of the earlier two phases. 

\begin{eqnarray*}
 && f(t) = \cos(\omega_\alpha t + \theta_\alpha) + \cos(\omega_\beta t + \theta_\beta) + \cos(\omega_\gamma t + \theta_\gamma) \\
 && \frac{1}{\omega_\gamma} = \frac{1}{\omega_\alpha} + \frac{1}{\omega_\beta}, 
\end{eqnarray*}
where  $\theta_\alpha, \theta_\beta,$ and $\theta_\gamma$ are all random phases.

In the other copy of the dataset, the study will add a sinusoidal dataset of superposition of the two frequencies and a third phase, which is the addition of the earlier two phases. 

\begin{eqnarray*}
&& f(t) = \cos(\omega_\alpha t + \theta_\alpha) + \cos(\omega_\beta t + \theta_\beta) + \cos(\omega_\gamma t + \theta_\gamma) \\
&& \frac{1}{\omega_\gamma} = \frac{1}{\omega_\alpha} + \frac{1}{\omega_\beta}, ~\text{and} ~\theta_\gamma = \theta_\alpha + \theta_\beta
\end{eqnarray*}
\noindent where $\theta_\alpha$, $\theta_\beta$ are random phases.
Now, the study will plant some low-amplitude artificial noise in both datasets. However, this addition of noise is optional and it only indicates the robustness of the diagnostic technique we follow. Thus, both the illustrative sample data consist of an amalgamation of three different cycles, which is hard to distinguish by a quick specular inspection. The first dataset does not possess any fixed phase relation between the different cycles, while the second one, by construction, contains a phase relationship between the three modes. Now let us take the Fourier transform of these two datasets. The Fourier transform will provide amplitudes in the complex number domain. To draw any meaningful conclusion, we will have to calculate the modulus value (the square norm) of the amplitudes. This will indicate the strength of each of the cycles in the data. However, this also throws away the phase relations that were otherwise present in the complex amplitudes. Thus, a regular Fourier transform of the two datasets will look identical and the information that the superposed frequency had a pre-specified phase relationship in the second dataset can never be retrieved from this analysis. 
Here in this study, we propose an extended form of integral transform (to be specific, an extended Fourier transform) that is capable of retaining this phase relationship. This method has been explored earlier in the context of electrical engineering and some turbulence-related studies in astrophysics as well as plasmas in nuclear fusion reactors. The method is simple to implement and easy to evaluate. First, the study will calculate the complex conjugate of the Fourier modes. Then multiply two different complex Fourier modes and further multiply them with the complex conjugated Fourier mode where the mode number is the addition of the previous two modes. This study will further sum over all such possible combinations. This new quantity retains the phase information of all the modes. This can be easily tested from the two illustrative datasets that was prepared earlier. The first dataset will not show any correlation, while the second one will now show spikes for those modes which has been used to create the third mode. (Refer to Appendix)
\subsection{Extended Fourier Transform}
In this study, we will construct a new quantity as 
\begin{equation}
    p(\omega_\alpha)=F(\omega_\alpha)F(\omega_\beta)F^*(\omega_\alpha+\omega_\beta)
\end{equation}
Here we multiply two different Fourier components $F(\omega_\alpha)$ and $F(\omega_\beta)$ and then multiply the result with the complex conjugate of Fourier modes at frequency $(\omega_\alpha+\omega_\beta)$ as previously mentioned in \citep{sharma2025}.\\

By undergoing the complex conjugate of the Fourier modes when FT is carried out a function (wave or signal) we express it as a sum of sinusoidal components (wave) of different frequencies these components are Fourier modes) we are changing the direction of the wave to retain the phase information.\\

Multiplying the two complex Fourier modes  $F(\omega_\alpha)$ and $F(\omega_\beta)$ represents the interaction between the frequency component $\omega_\alpha$ and $\omega_\beta)$. When we multiply the two complex modes with the complex conjugate mode, $F(\omega_\alpha+\omega_\beta)$, this will ensure that the new quantity $P(\omega_\alpha,\omega_\beta)$ retains the phase information or connections of interacting modes. If strong phase interaction exists between $\omega_\alpha$ and $\omega_\beta$ it will appear as strong correlation (sharp peak) in $P(\omega_\alpha,\omega_\beta)$.\\

Let us consider this extended Fourier transform in the earlier two data sets that we have prepared. The 1st data set (shows no correlation), which means if there is no strong interaction between different modes then $P(\omega_\alpha,\omega_\beta)$  will not show significant values which means that mode interactions are weak or do not exist. 2nd data set (shows correlated modes), which means that if certain modes interact to form the third mode,i.e. 
$F(\omega_\alpha)$ and $F(\omega_\beta)$contribute to  $F(\omega_\alpha,\omega_\beta)$, then $P(\omega_\alpha, \omega_\beta)$ will show peaks for those interactions which is because the phase relations are preserved (Refer Appendix). At the beginning this study uses simulated data sets of two different situations. First, where the all their modes are independent and second, where the third mode is made up of addition of earlier two phases and tested this developed simulated data in the model. We choose $\omega_\alpha=0.22$ and $\omega_\beta= 0.375$ for our simulated data.
\begin{figure}[h]
  \centering
  \begin{minipage}{0.3\textwidth}
    \centering
    \includegraphics[width=\linewidth]{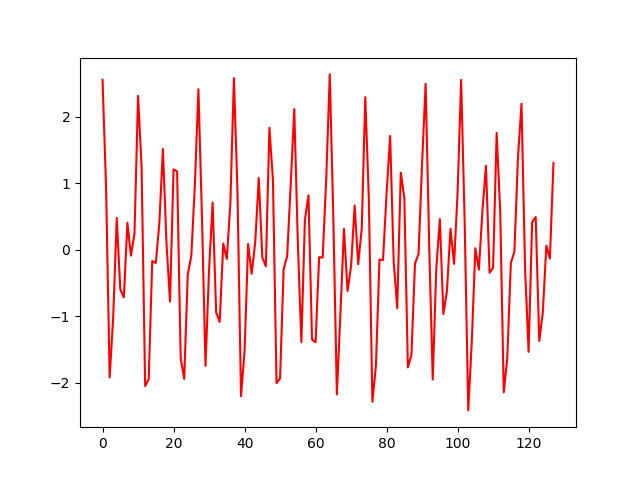}
  \end{minipage}
  \begin{minipage}{0.3\textwidth}
    \centering
    \includegraphics[width=\linewidth]{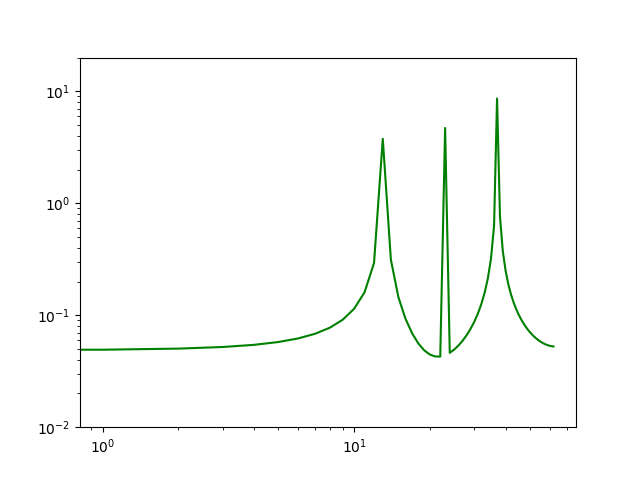}
  \end{minipage}
  \begin{minipage}{0.3\textwidth}
    \centering
    \includegraphics[width=\linewidth]{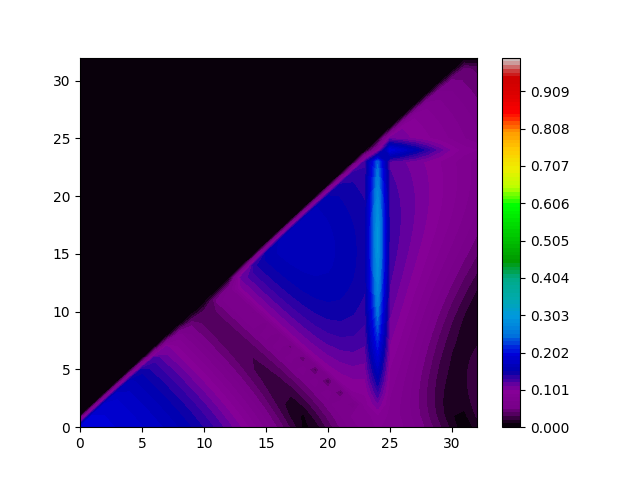}
  \end{minipage}
      \caption{Raw data, spectrum and the extended integral transform of the second set of artificially generated data without correlation.}
      \label{Data2}
\end{figure}

\begin{figure}[h]
  \centering
  \begin{minipage}{0.3\textwidth}
    \centering
    \includegraphics[width=\linewidth]{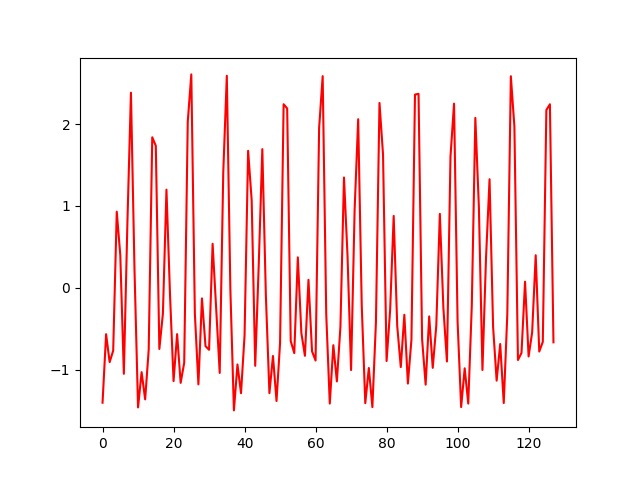}
  \end{minipage}
  \begin{minipage}{0.3\textwidth}
    \centering
    \includegraphics[width=\linewidth]{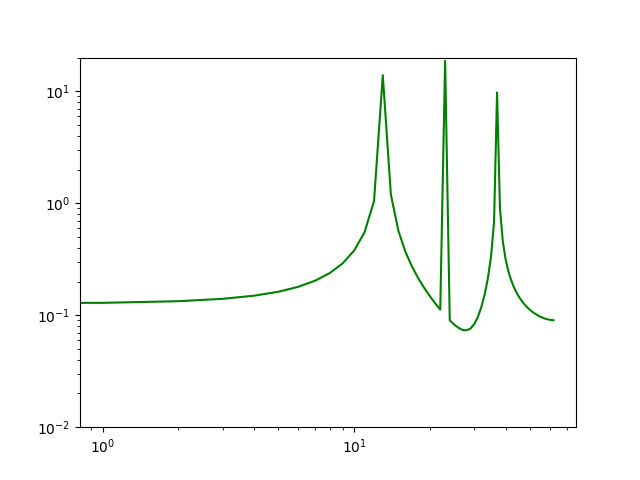}
  \end{minipage}
  \begin{minipage}{0.3\textwidth}
    \centering
    \includegraphics[width=\linewidth]{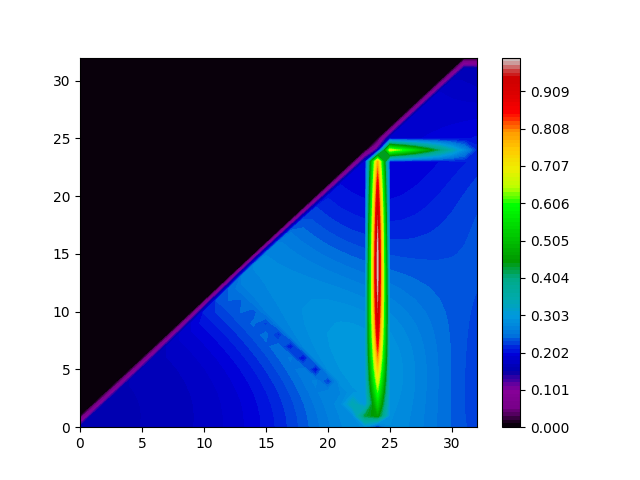}
  \end{minipage}
\caption{Raw data, spectrum and the extended integral transform of second set of artificially generated data with correlation.}
      \label{Data1}
\end{figure}
\subsection{Test of EFT on pure noise}
\begin{figure}[h]
  \centering
  \begin{minipage}{0.3\textwidth}
    \centering
    \includegraphics[width=\linewidth]{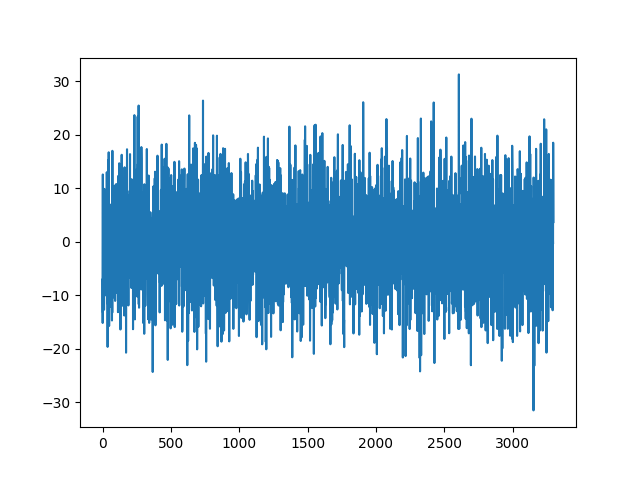}
  \end{minipage}
  \begin{minipage}{0.3\textwidth}
    \centering
    \includegraphics[width=\linewidth]{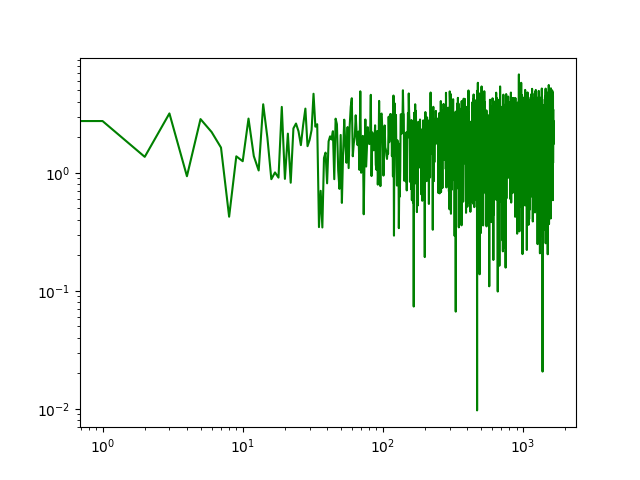}
  \end{minipage}
  \begin{minipage}{0.3\textwidth}
    \centering
    \includegraphics[width=\linewidth]{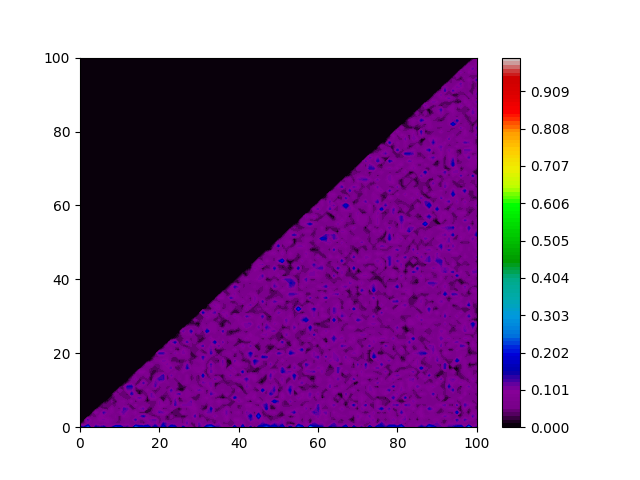}
  \end{minipage}
\caption{Uniform Random Numbers, their spectrum and the extended integral transform.}
      \label{URN}
\end{figure}

Before we employ our numerical method for the analysis of stock market data, we perform two tests. The first one is testing our Extended Fourier Transform (EFT) on a set of pure white noise, and the second one is the same on a colored noise. In the first case, we generate a sequence of 660000 white noise data using random number generator with zero mean and evaluate its Fourier transform. The EFT is also calculated as in Figure \ref{URN}. As expected, it shows no particular peak, since the white noise does not come with any particular phase correlation artificially planted inside it. Thus this signifies, there is no particular phase correlation is present in the whole dataset.

\begin{figure}[h]
  \centering
  \begin{minipage}{0.3\textwidth}
    \centering
    \includegraphics[width=\linewidth]{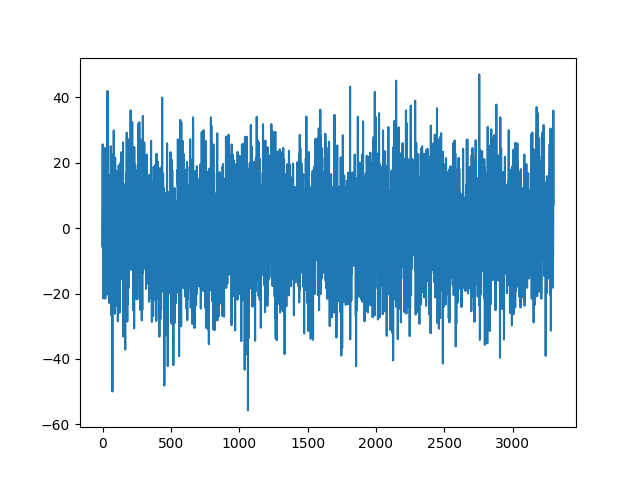}
  \end{minipage}
  \begin{minipage}{0.3\textwidth}
    \centering
    \includegraphics[width=\linewidth]{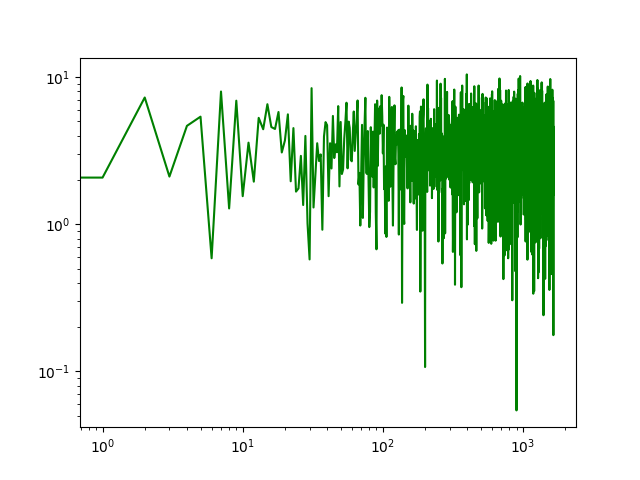}
  \end{minipage}
  \begin{minipage}{0.3\textwidth}
    \centering
    \includegraphics[width=\linewidth]{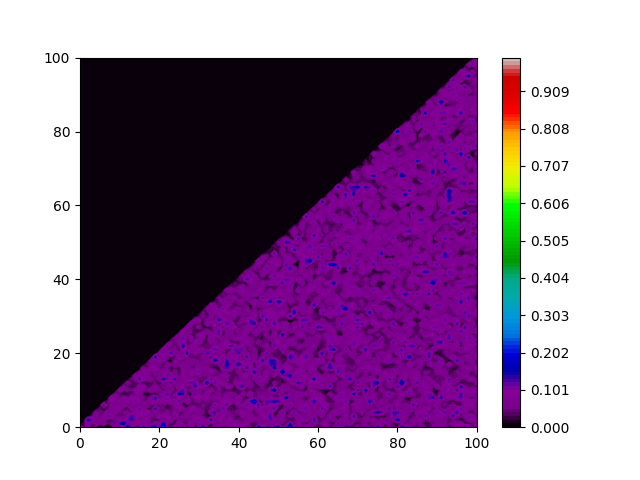}
  \end{minipage}
\caption{Gaussian Random Numbers, its spectrum and the extended integral transform.}
      \label{GRN}
\end{figure}

\begin{figure}[h]
  \centering
  \begin{minipage}{0.3\textwidth}
    \centering
    \includegraphics[width=\linewidth]{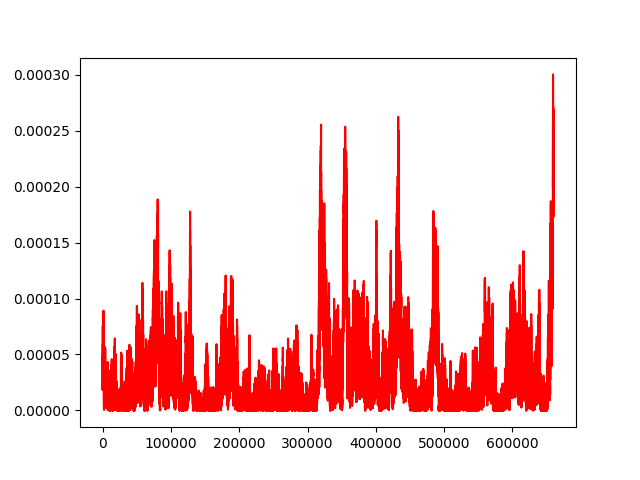}
  \end{minipage}
  \begin{minipage}{0.3\textwidth}
    \centering
    \includegraphics[width=\linewidth]{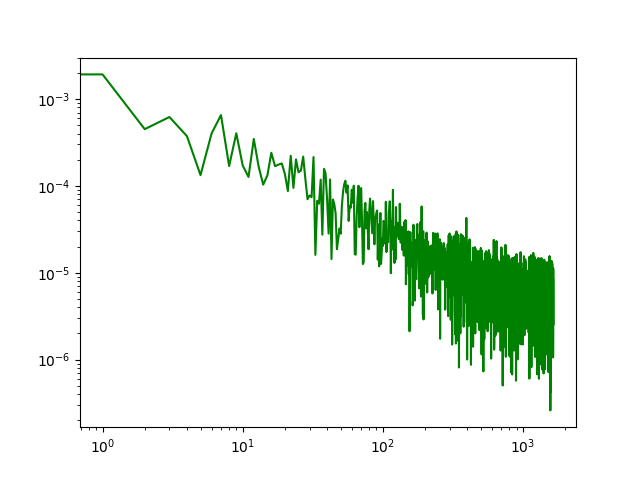}
  \end{minipage}
  \begin{minipage}{0.3\textwidth}
    \centering
    \includegraphics[width=\linewidth]{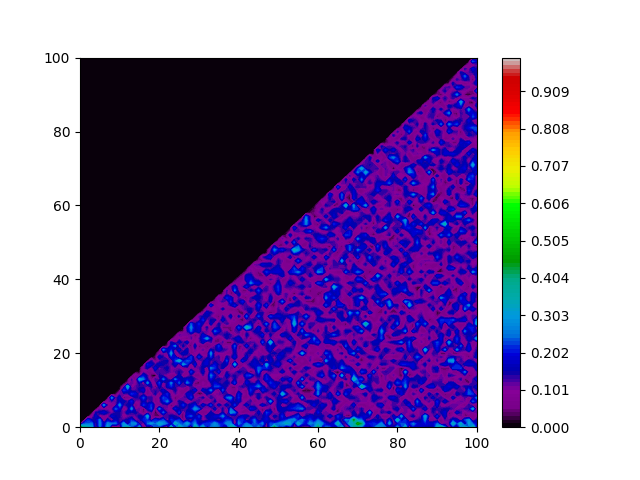}
  \end{minipage}
\caption{Simulated turbulence data with forcing and nonlinear interactions, spectrum and the extended integral transform.}
      \label{Burger}
\end{figure}

Further we evaluate the same for a colored noise. We generate the sample data of colored noise using the Box-Muller transform with two uniform random numbers ($U_1$ and $U_2$), both with zero mean. Further we evaluate the quantity $G = \sqrt{-2 \log U_1} \cos (2 \pi U_2)$. This provides us a sequence of Gaussian random number sequence with zero mean and standard deviation unity ($\mu = 0$ and $\sigma = 1$). This signifies a colored noise. We evaluate the Fourier transform of this time series data ($G$) and plot in Figure \ref{GRN}. The Fourier transform in log-log scale does not show any decay in the short scale. However this is not the case in fluid as well as the stock market turbulence data, as will be seen in the later sections.
Further, we test our Extended Fourier Transform (EFT) analysis on simulated turbulence data. We simulate Burgers equation 

\begin{eqnarray}
\frac{\partial u}{\partial t} + u \frac{\partial u}{\partial x} = \nu \frac{\partial^2 u}{\partial x^2} + f
\end{eqnarray}

using TARA computational framework described in \citep{mukherjee2019.41}, \citep{saikia2024.53}, \citep{mukherjee2019.45}. Here we use a spatial grid resolution of $N = 2^{10}$ and temporal resolution of $dt = 10^{-4}$. The simulation box is of length $L = 2 \pi$ and the initial condition is set as $u(0) = \sin(x)$. We choose $\nu = 3 \times 10^{-3}$ and the amplitude of forcing is set to $A = 6$. The external forcing is applied by a white noise with amplitude $A$ and zero mean. Since this represents an externally forced decaying turbulence, the energy spectra in large wavenumbers show a slight depreciation, as expected from any standard turbulence spectrum (see Figure \ref{Burger}).
Further we apply the same diagnostic tool for the evolution of a diffusion equation in presence of the same white noise type of forcing. The equation we solve is the following:

\begin{eqnarray}
\frac{\partial u}{\partial t} = \nu \frac{\partial^2 u}{\partial x^2} + f
\end{eqnarray}

\begin{figure}[h]
  \centering
  \begin{minipage}{0.3\textwidth}
    \centering
    \includegraphics[width=\linewidth]{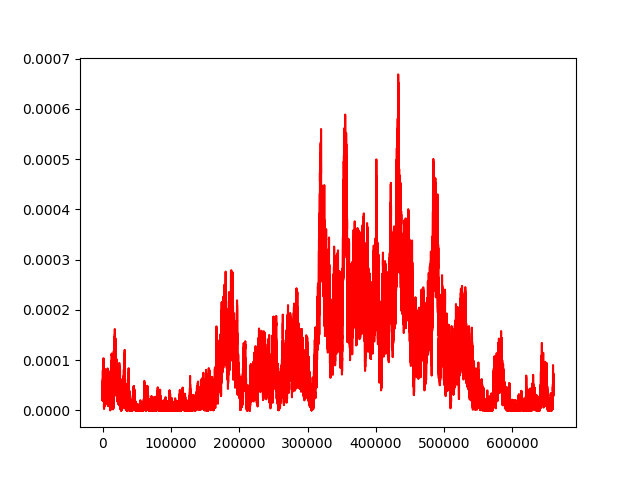}
  \end{minipage}
  \begin{minipage}{0.3\textwidth}
    \centering
    \includegraphics[width=\linewidth]{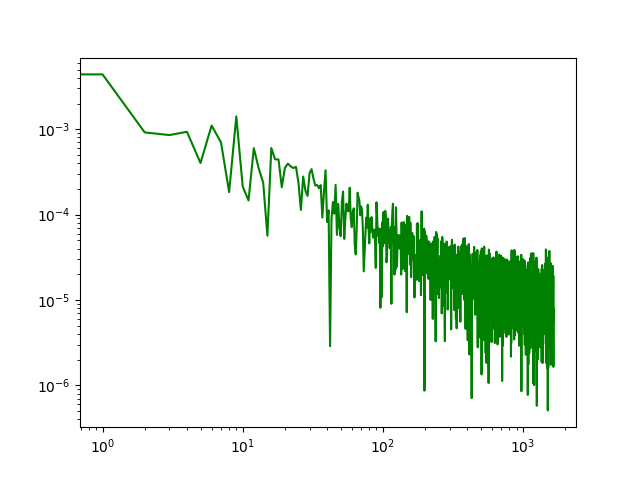}
  \end{minipage}
  \begin{minipage}{0.3\textwidth}
    \centering
    \includegraphics[width=\linewidth]{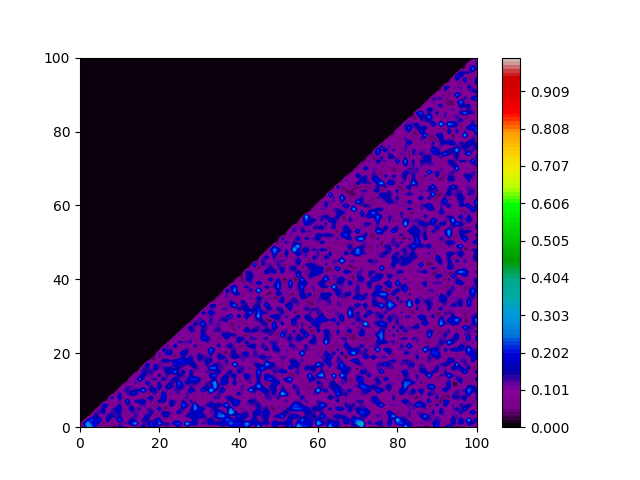}
  \end{minipage}
\caption{Simulated diffusive fluid turbulence data, spectrum, and the extended integral transform.}
      \label{Diffusion}
\end{figure}

We plot the raw data, the Fourier transform, and the EFT in Figure \ref{Diffusion}. All the above cases conclude that there is no phase correlation in the noisy turbulent dataset that we generated using our TARA simulation toolkit.
\subsection{Period of Study and Nature of Data}
 In this study we use secondary data from the stocks listed in NSE. The study uses tick data of 1-minute interval from 2015 to 2022, including the Nifty 50 Index data. The stock prices data consist of the stock prices of two major companies each from five different industries: Banking, Fast-Moving Consumer Goods, Automobiles, Information Technology, and Iron and Steel. This study uses unaltered or raw data from the above-mentioned companies under different industries so that a precise and accurate study can be conducted. Refer \ref{table1} 
\begin{center}
\begin{tabular}{|c|c|c|c|c|}
\hline
\textbf{Banking} & \textbf{FMCG} & \textbf{Automobiles} & \textbf{IT} & \textbf{Iron and Steel} \\
\hline
ICICI & Dabur & Mahindra and Mahindra & Infosys & JSW Sreel \\
\hline
SBI & HUL & Tata Motors & Tata Consultancy & Tata Steel \\
\hline
\label{table1}
\end{tabular}
\end{center}
\section{Results and Discussions}
Using the extended form of the Fourier transform to analyze two different simulated data sets reveals different outcomes when visualized. In the first data set, where all Fourier modes are independent, no spikes are observed, indicating a lack of correlation and interdependence between the diverse modes. Conversely, when the third mode is generated using the frequency components of the first two modes, spikes are observed, signaling a relationship and dependence between the modes. The concepts of frequencies and cycles is used to analyze the system of turbulence as discussed in earlier sections, in the state of fully developed turbulence it is expected to have no correlations or interactions between the phases and expected to have its own independent phase information and if there is dependence among the phases then it is concluded that the medium is not a fully developed turbulence. Using these concepts of phases for turbulence, it is found that in first data set there is no such spikes which indicates that the first data set has no relation between the phases and are totally independent and are turbulent in nature. On another data set where the third phase is generated through modes of different phases shows spikes indicating the inter dependence and correlation among phases. This situation of correlation indicates between the phases indicates that this medium is not a fully developed turbulence (Refer to \ref{Data2} \ref{Data1}).

\begin{figure}[h]
  \centering
  \begin{minipage}{0.3\textwidth}
    \centering
    \includegraphics[width=\linewidth]{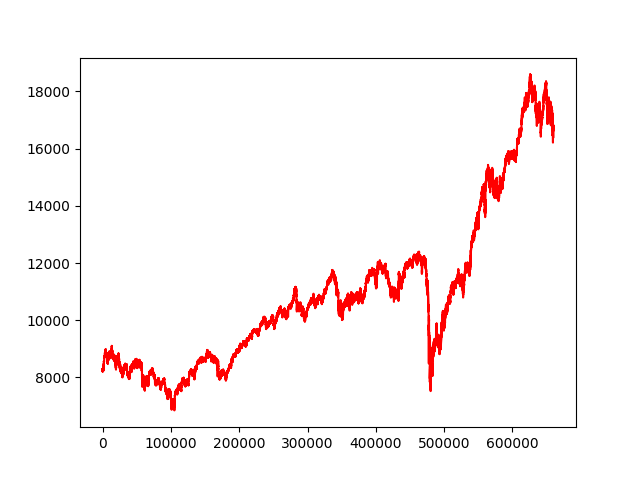}
  \end{minipage}
  \begin{minipage}{0.3\textwidth}
    \centering
    \includegraphics[width=\linewidth]{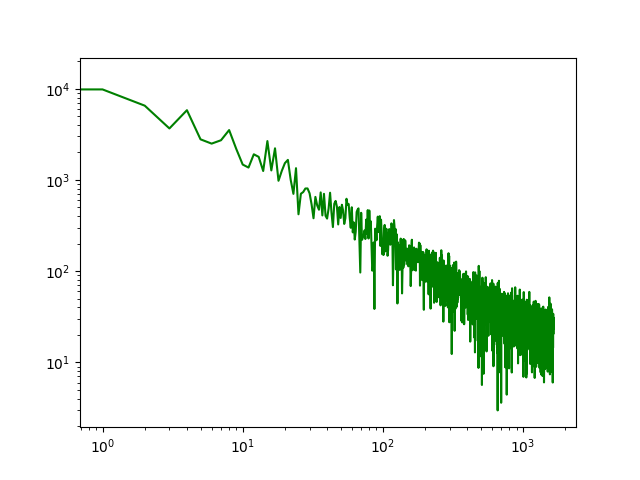}
  \end{minipage}
  \begin{minipage}{0.3\textwidth}
    \centering
    \includegraphics[width=\linewidth]{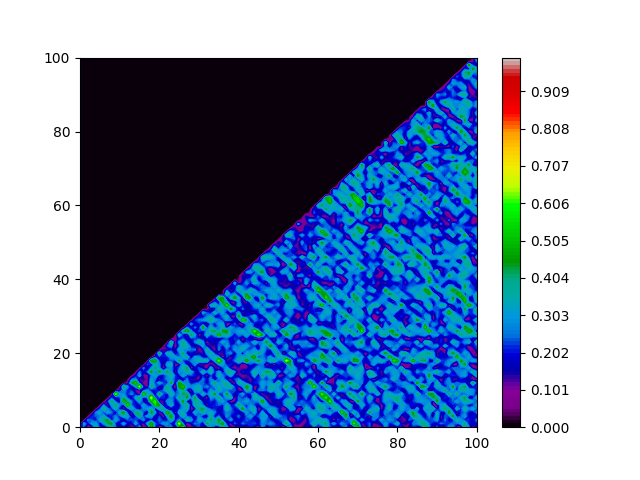}
  \end{minipage}
\caption{Raw data, Turbulence spectrum and the extended integral transform of Nifty 50 Index data.}
      \label{Nifty50}
\end{figure}

\begin{figure}[h]
  \centering
  \begin{minipage}{0.3\textwidth}
    \centering
    \includegraphics[width=\linewidth]{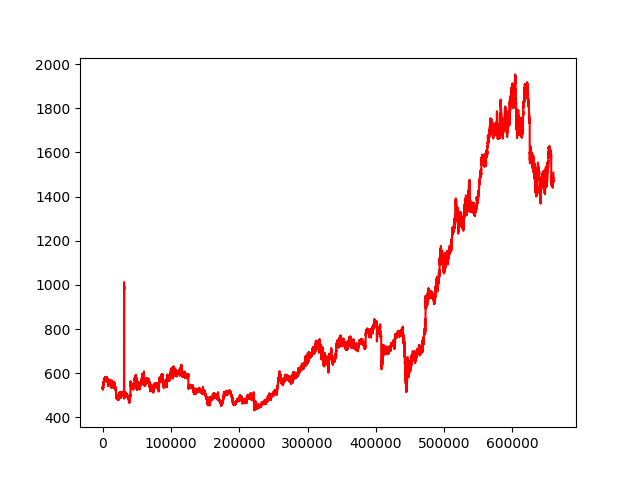}
  \end{minipage}
  \begin{minipage}{0.3\textwidth}
    \centering
    \includegraphics[width=\linewidth]{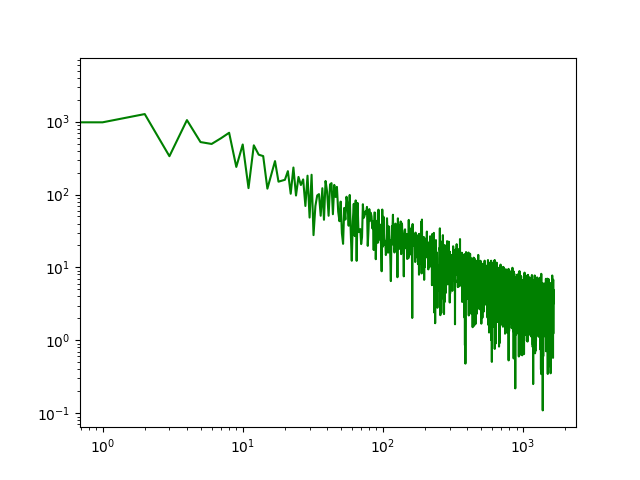}
  \end{minipage}
  \begin{minipage}{0.3\textwidth}
    \centering
    \includegraphics[width=\linewidth]{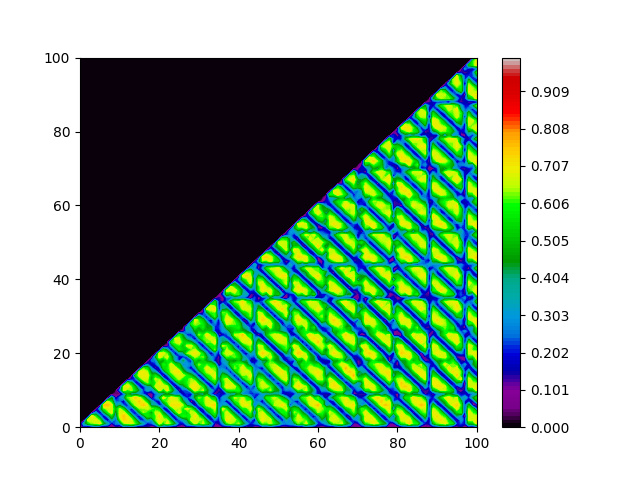}
  \end{minipage}
\caption{Raw data, Turbulence spectrum, and the extended integral transform of Infosys data.}
      \label{Infosys}
\end{figure}

This method of transitioning data from the time domain to the frequency domain and examining various frequency modes is essential for understanding the interdependence among different Fourier modes in stock market time series data. This information on frequency modes is used to understand the nature of the stock market as a fully developed turbulence or not. When this study applies these extended Fourier methods to understand the phase information and the nature of turbulence in the stock market data. Here we consider only two stock data one is Nifty 50 and another is INFOSYS \ref{Nifty50} \ref{Infosys}. For the output of other stocks, refer appendix.

While analyzing the outcomes of the turbulent spectrum of stock data, especially the second picture of each figure, it is very hard to differentiate between fully developed turbulence, as all of them look identical and resemble the qualitative nature of turbulence. but the quantitative nature of turbulence is not visible in the turbulence spectrum due For this reason, the extended integral transform analysis is used to identify the nature of the financial market. A 2-D plot of extended integral transform analysis is reported in this paper so that a clear and better understanding can be made in on quantifying the stock market turbulence.
 
Analyzing stock data from the above-mentioned perspectives yields a wealth of information, allowing us to draw numerous conclusions and assumptions. The second dataset provides insights into phase relations, exhibiting spikes as shown in Figure 3. A similar trend can be observed in the stock market data for Infosys, illustrated in Figure 9. These spikes indicate interactions and relationships between the various frequency modes when analyzed using the Extended Fourier Transform (EFT). However, in other stock datasets, we do not observe such interactions between the two modes to create a third mode. Our analysis of stock market data using the Extended Fourier Transform focused on retaining the phase relations among different Fourier modes. This method reveals spikes when visualized. In the case of Infosys, we identified spikes that signify an interplay within the Fourier modes of its data cycle. In contrast, other stock prices do not exhibit such relationships, as seen in the first dataset. Furthermore, the Infosys dataset shows results consistent with the second dataset, where the phases and modes are interdependent. This implies that the cyclical pattern of Infosys is not independent. It suggests that factors affecting the stock market may have influenced the price fluctuations in Infosys. This raises the possibility that information may have been artificially introduced into the price of Infosys stock either publicly or privately leading to this observed behavior.

Considering the concepts of phases and their relation with the turbulence, is expected that in the state of fully developed turbulence, there are no mutual dependence among the phases, and if there are no such relations, the phases are independent, concluding that it is not a fully developed turbulence. Considering this concept of turbulence to understand the financial market and the phase information of time series data when converted to frequency domain reveals that the modes of Infosys are not independent and is influenced by other phases. whenever there is some dependence among the phases then the spikes appear signaling that Infosys is not turbulent in nature. But in the cases of other stock prices data, it has been revealed that there is correlation among the modes which implies that stocks exhibits the state of fully developed turbulence which is also visible through the extended integral transform analysis. The other nine stock prices of different companies across five different industries and the Nifty 50 Index data of the National Stock Exchange of India go through a state of fully developed turbulence. The Nifty 50 index data also shows a nature of turbulence, which means there must be turbulence in the NSE, but surprisingly, Infosys shows a different nature, as not to be fully developed turbulence. This situation in the Infosys stock market reveals that there may be certain information related to the market of Infosys that causes such unusual behavior of the prices of Infosys, or it may have been influenced by some environment, as well as another market factor which caused such distinct behavior.
\section{Conclusion}
The analysis of the financial market through a computational approach is capable of yielding better results and provides a deeper and unique approach to analyze the financial data. In this article, we employ theories and concepts of mathematical physics using computational physics, specifically the Fourier transform, which is commonly used to study the turbulence and phase properties of physical time series data. This study proposes a mathematical model called the Extended Fourier Transform (EFT). At first, we tested the effectiveness of this model on two different sample datasets and then extended its application to stock market data, which included ten stock prices and the Nifty 50 index. When the proposed model is applied to the draw interface of the data set, it is very difficult to draw conclusions only through the turbulence spectrum, so this study proposes an extended form of the integral transform to understand the underlying nature of turbulence in a deeper sense. The extended Fourier transform is capable of retaining the phase information between the modes, and spikes will appear when visualized through the bi-spectrum.  While using this proposed model in the sample data set, it was found that there is some correlation between the modes in the second sample data set, as this data set showed spikes when visualized, but no such inference can be drawn from the other sample data set. Considering the phase properties, it is to be noted that in a state of fully developed turbulence, there are no correlations among the phases, and if there are some associations between the modes or phases, then such a state cannot be considered a fully developed turbulence. The same has been observed in the first data set, and this data set is considered to be a state of fully developed turbulence. The same extended Fourier Transform, which is capable of retaining the phase relations, has been applied to the stock market. When the frequency domain of the Infosys prices was analyzed, it was found that the phases which was decomposed from its data revealed some correlation between them.. Further, it is considered that the market of Infosys is not fully developed state of turbulence. However, the results for the other stock market data were different and independent modes were present, showing no such correlation, concluding to have a fully developed turbulence in these stocks, including the Nifty 50 index data. The stock prices of all the major stocks, along with the Nifty 50 index data, showed the nature of fully developed turbulence, but Infosys displayed no sign of fully developed turbulence. This different behavior of Infosys implies that the prices of Infosys may have been influenced by some environmental as well as market factors. This study contributes to the field of econophysics by applying the theories and methods of computational physics to model the financial market turbulence. Furthermore, this study also aims to enhance the stakeholders' knowledge to understand the nature of the market through frequency and phase analysis. 

\bibliography{reference}
\section{Appendix}
\subsection{FOURIER TRANSFORM}
\begin{equation*}
    F(\omega) = \frac{1}{\sqrt{2\pi}} \int_{-\infty}^{\infty} f(t) \, e^{i\omega t} \, dt
\end{equation*}
Here $f(t)=$original function/signal
$F(\omega)=$ Result of Fourier Transform which tells us how much of each frequency $(\omega)$ is present in the signal.$e^{i\omega t}$ is the way of representing waves (it combines sine and cosine waves using complex numbers$(a+\iota b$,where a and b are real and $\iota$ is imaginary unit)

This equation helps us to see "ingredients" Frequencies(how fast the wave comes.example 1HZ= 1 wave per second/how many wave pass a point in second) that make the signal. example if a signal consist of a mix of fast and slow waves the FT will show show fast and slow.
\subsection{DISCRETE FOURIER TRANSFORM(DFT)}
\begin{equation*}
    F(\omega) = \sum_{t=0}^{N-1} f(t) \cdot e^{-i \cdot \frac{2\pi \omega t}{N}}
\end{equation*}
$F(t)$ is our data at real time(stock price) tells us how much of each frequency$(\omega)$ is present in the data. The amplitudes(height of the wave) of the Fourier modes(FT expresses function as a sum of sinusoids with different frequencies, amplitudes, and phases). These individual sinusoids are the Fourier modes, which signify the strength of the corresponding frequency that is present in the time series data. Each one represents the cyclical pattern of data movement. when there are multiple frequencies present in the data, the cyclical movement becomes unmanageable, so it requires an in-depth analysis. Hence, we use FFTW to evaluate the FT of both the illustrative and stock market data.

When we use DFT on stock data, it generates Fourier modes in the complex number domain. We use the square norm/ modules of the complex data to draw conclusions, but the square norm, which maps the data to the real number, throws away the phase relation (where the wave starts), which, as a result, may lead to losing the important information hidden inside the data.

This study focuses on providing an extension to this transformation (the Fourier Transformation) such that the phase relations are retained.
\subsection{DETAILED TOOLS AND TECHNIQUES}
First we create a sinusoidal data set of fixed frequency and fixed phases $f(t)=cos(\omega_\alpha t+\theta_\alpha)$ we add another sinusoidal data set of fixed but different frequency and fixed phase other than the previous one.$cos(\omega_\beta t+\theta_\beta)$ which will be $f(t)=cos(\omega_\alpha t+\theta_\alpha)+cos(\omega_\beta t+\theta_\beta)$ and further we will create two copies of the same data set.\\

1st Data Set: which has no phase relations and $\theta_\alpha,
\theta_\beta,\theta_\gamma$
are all random and independent phases 

we will add a third sinusoidal wave with frequency $\omega_\gamma$ and an independent random phase $\theta_\gamma$ and it is represented as 

$f(t)=cos(\omega_\alpha t+\theta_\alpha)+cos(\omega_\beta t+\theta_\beta+cos(\omega_\gamma+\theta_\gamma)$

where $\omega_\gamma$ follows the relation $ \frac{1}{\omega_\gamma} = \frac{1}{\omega_\alpha} + \frac{1}{\omega_\beta},$((NOTE: phase $\theta_\gamma$ is independent of $\theta_\alpha$ and $\theta_\beta$ phases))\\

2nd Data Set: Fixed phase relation which means $\theta_\gamma=\theta_\alpha+\theta_\beta$

Instead of choosing an independent phase $\theta_\gamma$, we propose a fixed phase relation that is $\theta_\gamma=\theta_\alpha+\theta_\beta$ and the second data set is represented below

$f(t)=cos(\omega_\alpha t+\theta_\alpha)+cos(\omega_\beta t+\theta_\beta+cos(\omega_\gamma+\theta_\gamma)$
where $ \frac{1}{\omega_\gamma} = \frac{1}{\omega_\alpha} + \frac{1}{\omega_\beta}$,$\theta_\gamma=\theta_\alpha+\theta_\beta$ (NOTE: phase $\theta_\gamma$ is not independent.

The above two data sets consist of a joint nature of three different cycles and are very hard to easily distinguish these three cycles as the first data set does not have any fixed phase relation, but the second one has fixed phase relations. Both datasets are represented below.
\begin{figure}[h]
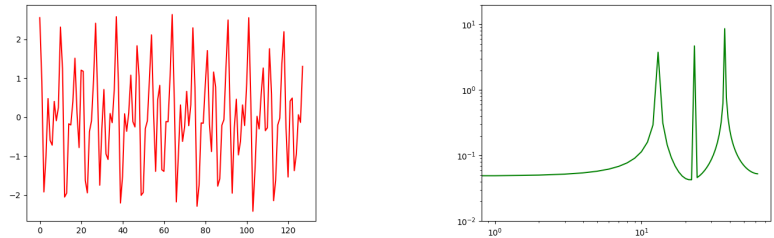

  \centering
  \begin{minipage}{0.3\textwidth}
    \centering
    \includegraphics[width=\linewidth]{data2.png}
  \end{minipage}
  \hspace{0.05\textwidth}
  \begin{minipage}{0.3\textwidth}
    \centering
    \includegraphics[width=\linewidth]{Freq_2.png}
  \end{minipage}
\caption{Raw data of the first and second sets of artificially generated data mentioned above.}
\label{raw_data}
\end{figure}
We will further take the Fourier transform of the given data sets. the Fourier transform will provide amplitudes(height of the wave/maximum value of the wave) in the complex number domain.
\begin{figure}[h]
  \centering
  \begin{minipage}{0.3\textwidth}
    \centering
    \includegraphics[width=\linewidth]{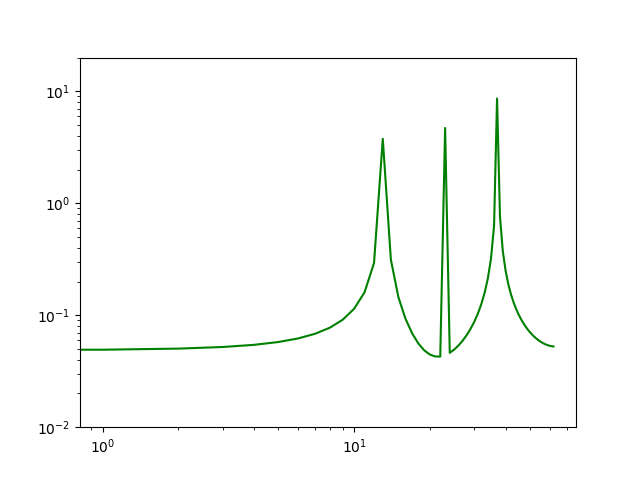}
  \end{minipage}
  \hspace{0.05\textwidth}
  \begin{minipage}{0.3\textwidth}
    \centering
    \includegraphics[width=\linewidth]{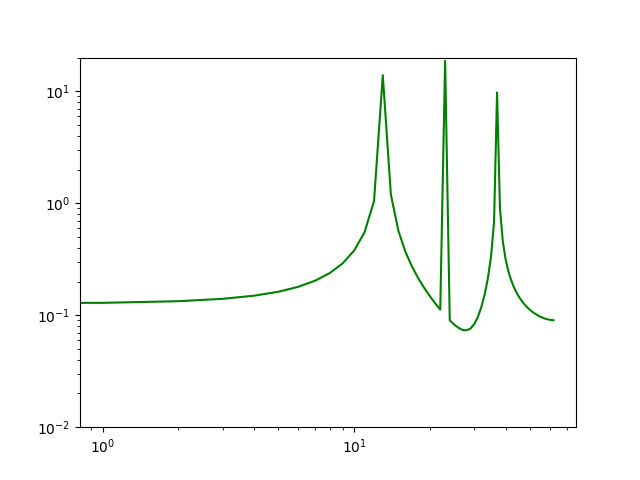}
  \end{minipage}
\caption{Frequency Spectrum of the first and second sets of artificially generated data.}
\label{raw_data}
\end{figure}
While applying the Fourier transform to the datasets, it is very difficult to distinguish between them as they are identical. Applying the square norm( tells the strength of cycles/frequency) discards the phase information which are present in the complex amplitude. For this reason, both datasets will look identical, even if there are specified phase relations. so we propose an Extended form of the Fourier Transform, which is capable of retaining the phase information.
\subsection{EXTENDED FOURIER TRANSFORM}
In this study, we will construct a new quantity as 
\begin{equation}
    p(\omega_\alpha,\omega_\beta)=F(\omega_\alpha)F(\omega_\beta)F^*(\omega_\alpha+\omega_\beta)
\end{equation}
Here we multiply two different Fourier components $F(\omega_\alpha)$ and $F(\omega_\beta)$ and then multiply the result by the complex conjugate of the Fourier modes at the frequency. By undergoing the complex conjugate of the Fourier modes (when FT is carried out on a function (wave or signal), we express it as a sum of sinusoidal components (wave) of different frequencies, these components are Fourier modes) we are changing the direction of the wave to retain the phase information. Multiplying the two Fourier modes $F(\omega_\alpha)$ and $F(\omega_\beta)$ represents the interaction between the frequency components $\omega_\alpha$ and $\omega_\beta$.

when we multiply the two modes with the complex conjugate $F^*(\omega_\alpha+\omega_\beta)$, this will ensure that the new quantity $P(\omega_\alpha,\omega_\beta)$ retains the phase information or connections of interacting modes. If a strong interaction exists between $\omega_\alpha$ and $\omega_\beta$ it will appear as a strong correlation in $P(\omega_\alpha,\omega_\beta)$.
Let us consider this extended Fourier Transform in the earlier two data sets that we have prepared 

The First data set shows no correlation, which means that if there is no strong correlation between modes, then $P(\omega_\alpha,\omega_\beta)$ will not show significant values, which means that the mode interactions are weak or do not exist.

On the Second data set which has correlated modes signifies that if a certain mode interacts to form the another mode, i.e $F(\omega_\alpha)$ and $F(\omega_\beta)$ contribute to $F(\omega_\alpha,\omega_\beta)$ then, $P(\omega_\alpha,\omega_\beta)$ will show peaks for the interaction due the phase relations are preserved.

\begin{figure}[h]
  \centering
  \begin{minipage}{0.3\textwidth}
    \centering
    \includegraphics[width=\linewidth]{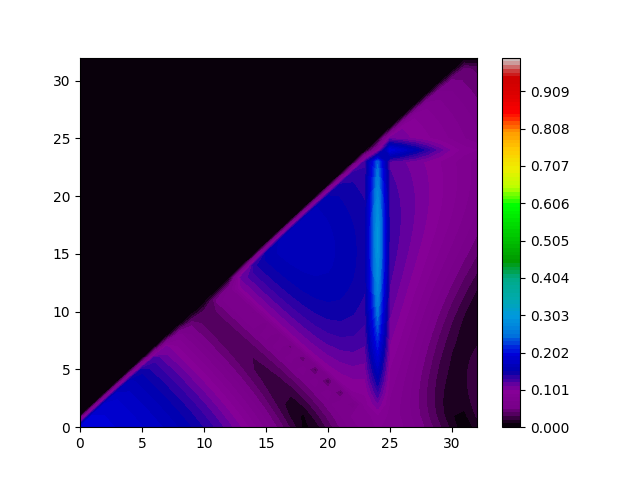}
  \end{minipage}
  \hspace{0.05\textwidth}
  \begin{minipage}{0.3\textwidth}
    \centering
    \includegraphics[width=\linewidth]{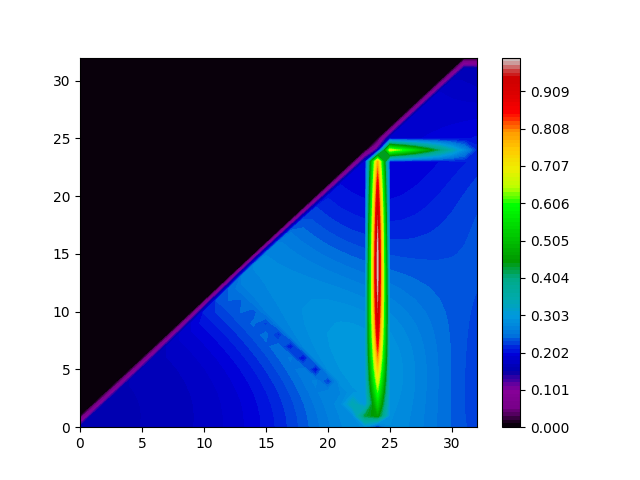}
  \end{minipage}
\caption{Extended integral transform of the first and second set of artificially generated data.}
\label{EIT}
\end{figure}

\subsection{ Raw Data, Turbulence Spectrum and Extended Integral transform of the Stock data of the above-mentioned companies of NSE.}

\begin{figure}[h]
  \centering
  \begin{minipage}{0.3\textwidth}
    \centering
    \includegraphics[width=\linewidth]{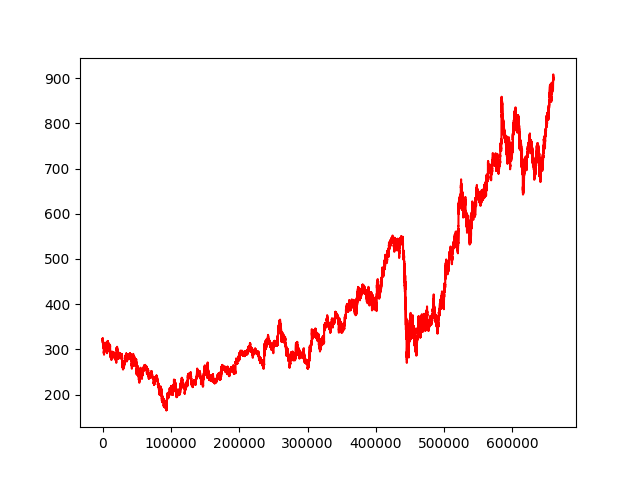}
  \end{minipage}
  \begin{minipage}{0.3\textwidth}
    \centering
    \includegraphics[width=\linewidth]{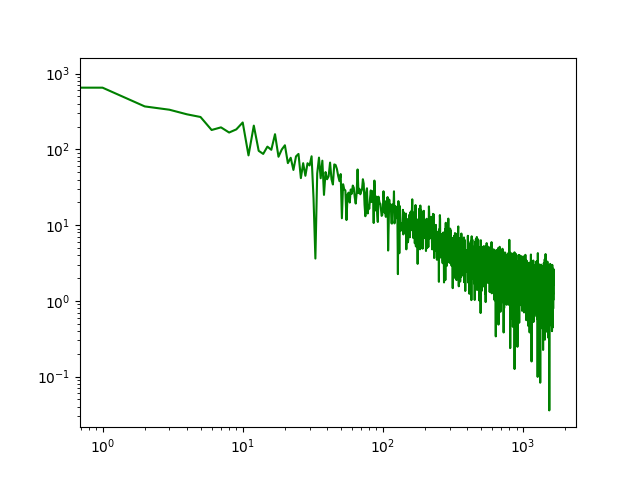}
  \end{minipage}
  \begin{minipage}{0.3\textwidth}
    \centering
    \includegraphics[width=\linewidth]{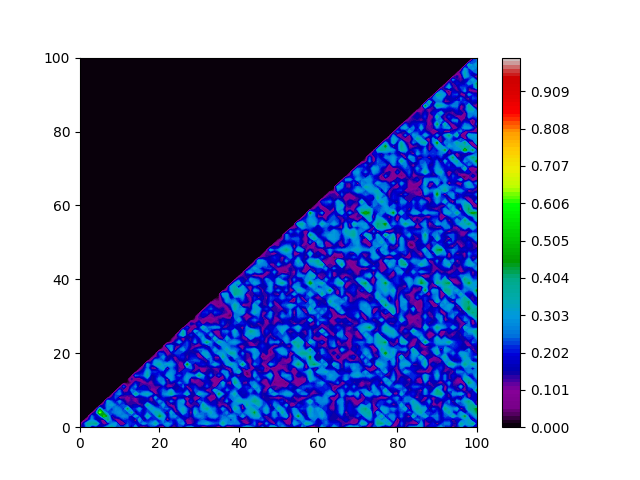}
  \end{minipage}
      \caption{Raw data, turbulence spectrum and the extended integral transform of ICICI Index data.}
      \label{ICICI}
\end{figure}

\begin{figure}[h]
  \centering
  \begin{minipage}{0.3\textwidth}
    \centering
    \includegraphics[width=\linewidth]{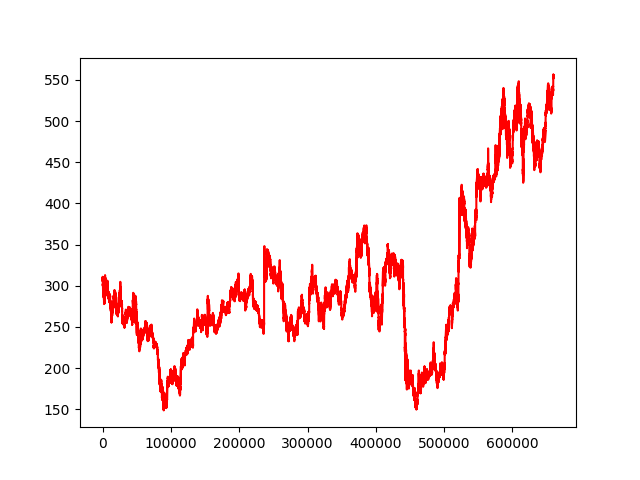}
  \end{minipage}
  \begin{minipage}{0.3\textwidth}
    \centering
    \includegraphics[width=\linewidth]{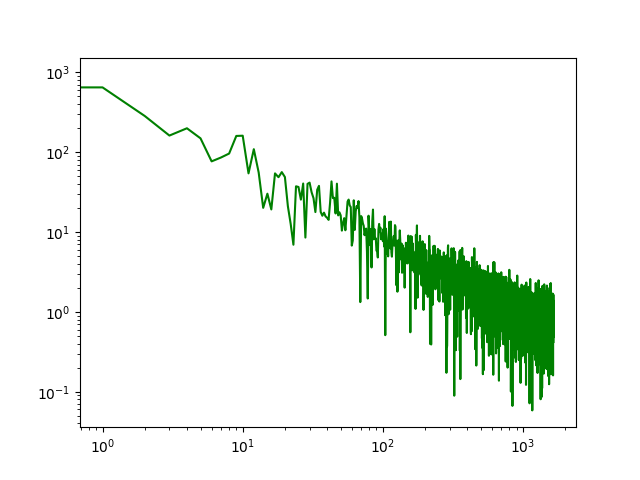}
  \end{minipage}
  \begin{minipage}{0.3\textwidth}
    \centering
    \includegraphics[width=\linewidth]{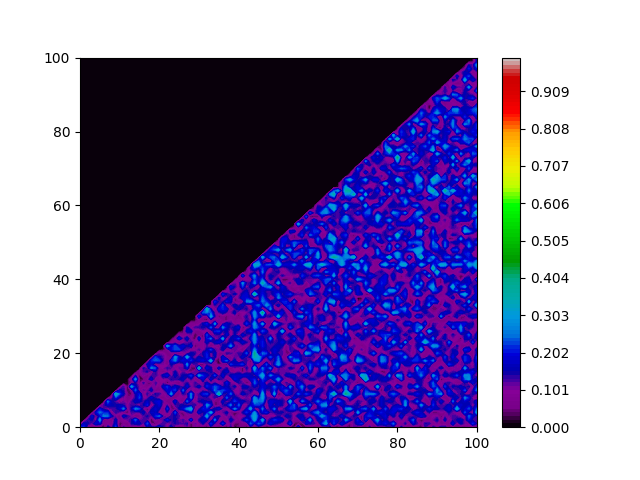}
  \end{minipage}
\caption{Raw data, turbulence spectrum and the extended integral transform of SBI Index data.}
      \label{SBI}
\end{figure}

\begin{figure}[h]
  \centering
  \begin{minipage}{0.3\textwidth}
    \centering
    \includegraphics[width=\linewidth]{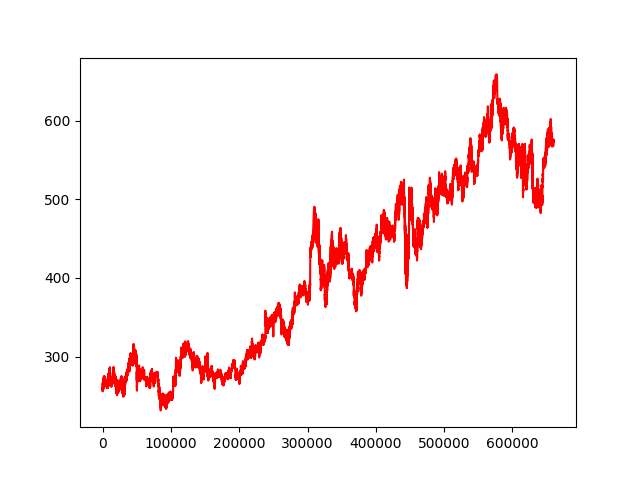}
  \end{minipage}
  \begin{minipage}{0.3\textwidth}
    \centering
    \includegraphics[width=\linewidth]{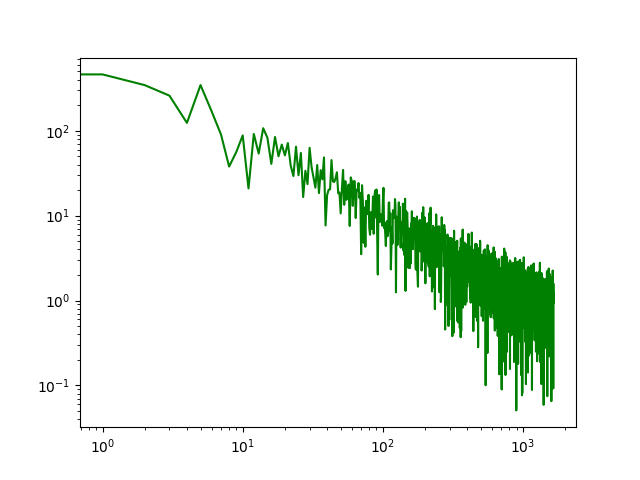}
  \end{minipage}
  \begin{minipage}{0.3\textwidth}
    \centering
    \includegraphics[width=\linewidth]{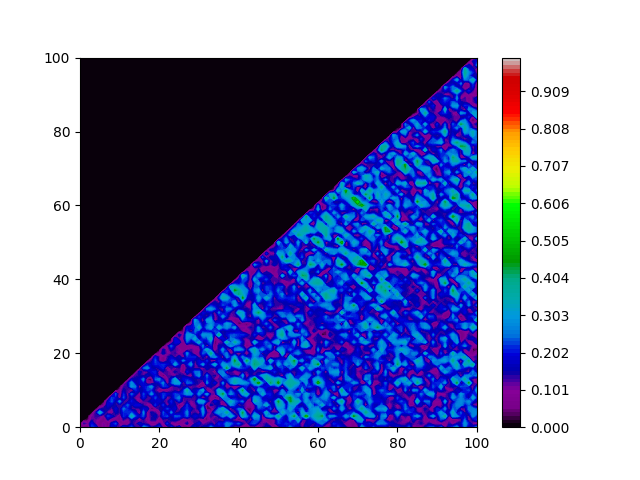}
  \end{minipage}
      \caption{Raw data, turbulence spectrum and the extended integral transform of Dabur Index data.}
      \label{Dabur}
\end{figure}

\begin{figure}[h]
  \centering
  \begin{minipage}{0.3\textwidth}
    \centering
    \includegraphics[width=\linewidth]{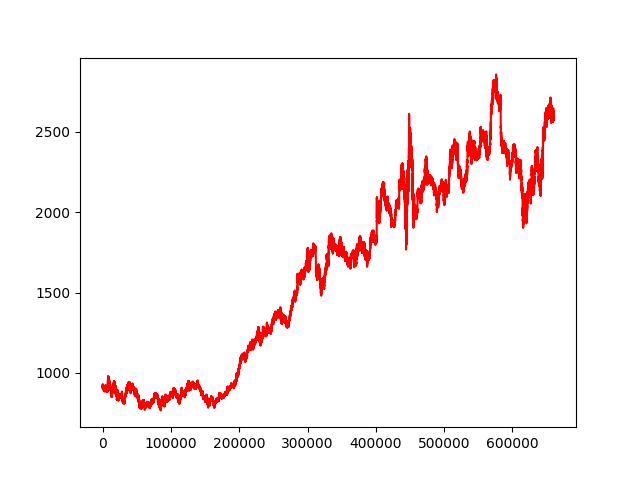}
  \end{minipage}
  \begin{minipage}{0.3\textwidth}
    \centering
    \includegraphics[width=\linewidth]{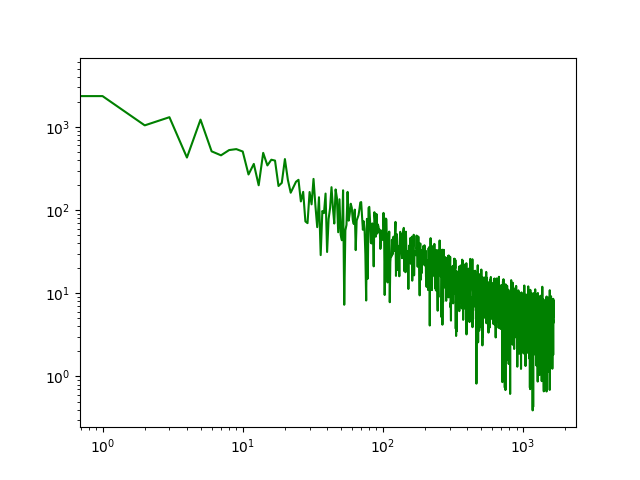}
  \end{minipage}
  \begin{minipage}{0.3\textwidth}
    \centering
    \includegraphics[width=\linewidth]{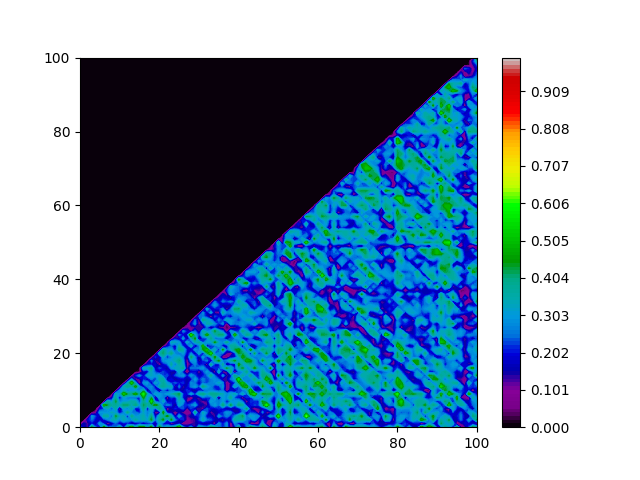}
  \end{minipage}
      \caption{Raw data, turbulence spectrum and the extended integral transform of HUL Index data.}
      \label{HUL}
\end{figure}

\begin{figure}[h]
  \centering
  \begin{minipage}{0.3\textwidth}
    \centering
    \includegraphics[width=\linewidth]{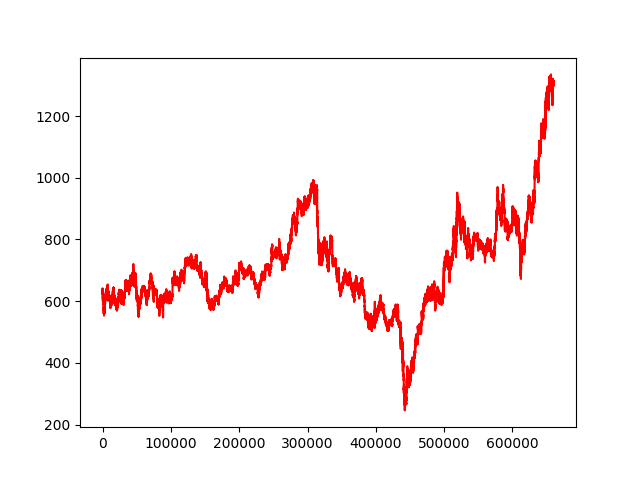}
  \end{minipage}
  \begin{minipage}{0.3\textwidth}
    \centering
    \includegraphics[width=\linewidth]{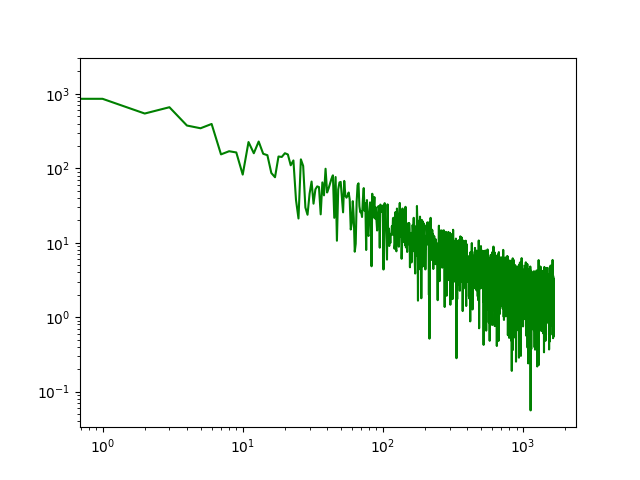}
  \end{minipage}
  \begin{minipage}{0.3\textwidth}
    \centering
    \includegraphics[width=\linewidth]{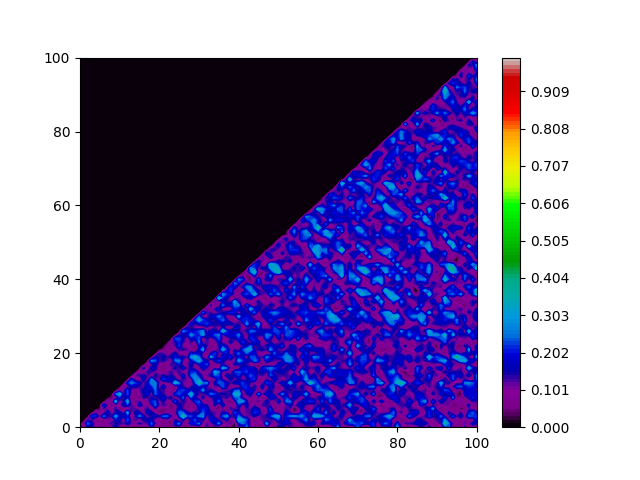}
  \end{minipage}
      \caption{Raw data, turbulence spectrum and the extended integral transform of M\&M Index data.}
    \label{MM}  
\end{figure}

\begin{figure}[h]
  \centering
  \begin{minipage}{0.3\textwidth}
    \centering
    \includegraphics[width=\linewidth]{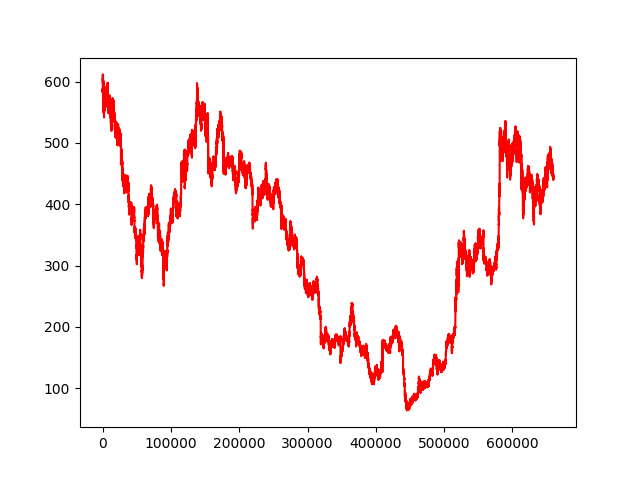}
  \end{minipage}
  \begin{minipage}{0.3\textwidth}
    \centering
    \includegraphics[width=\linewidth]{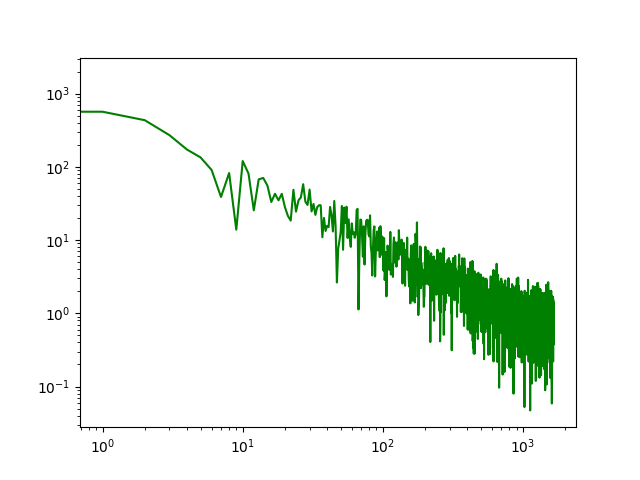}
  \end{minipage}
  \begin{minipage}{0.3\textwidth}
    \centering
    \includegraphics[width=\linewidth]{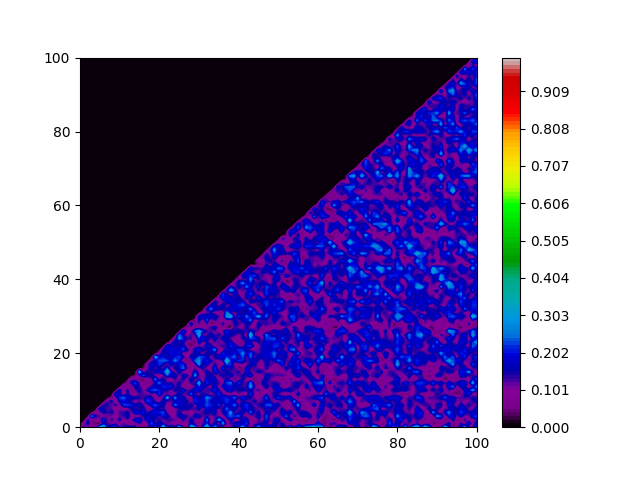}
  \end{minipage}
 \caption{Raw data, turbulence spectrum and the extended integral transform of Tata Motors Index data.} 
  \label{Tata_Motors}
\end{figure}

\begin{figure}[h]
  \centering
  \begin{minipage}{0.3\textwidth}
    \centering
    \includegraphics[width=\linewidth]{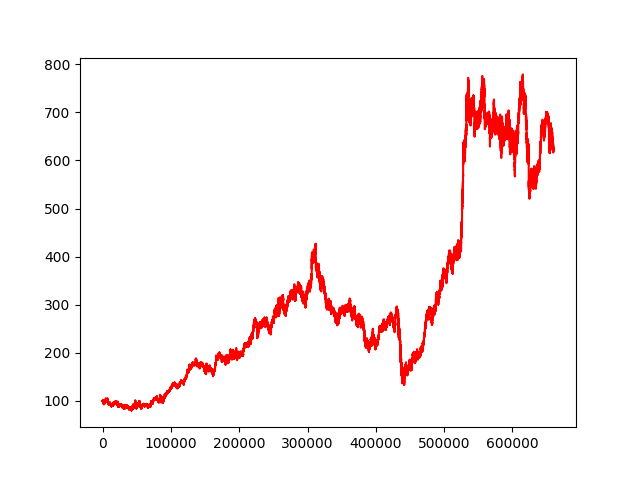}
  \end{minipage}
  \begin{minipage}{0.3\textwidth}
    \centering
    \includegraphics[width=\linewidth]{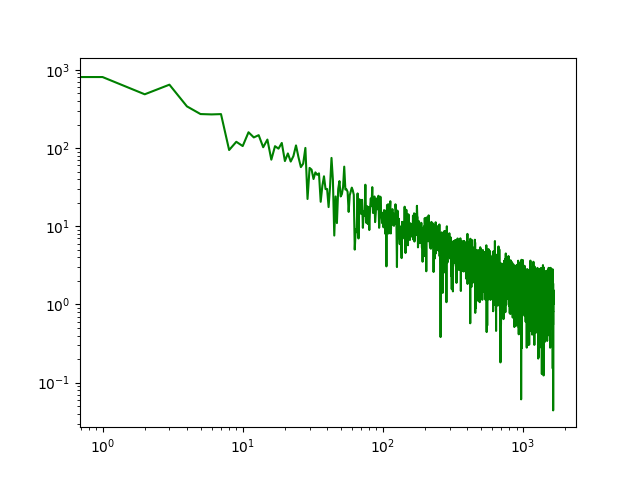}
  \end{minipage}
  \begin{minipage}{0.3\textwidth}
    \centering
    \includegraphics[width=\linewidth]{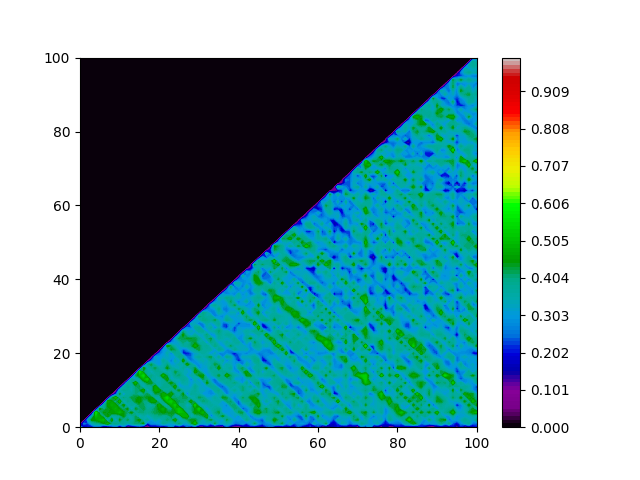}
  \end{minipage}
\caption{Raw data, turbulence spectrum and the extended integral transform of JSW Index data.}
\label{JSW}
\end{figure}

\begin{figure}[h]
\centering
    \begin{minipage}{0.3\textwidth}
    \centering
    \includegraphics[width=\linewidth]{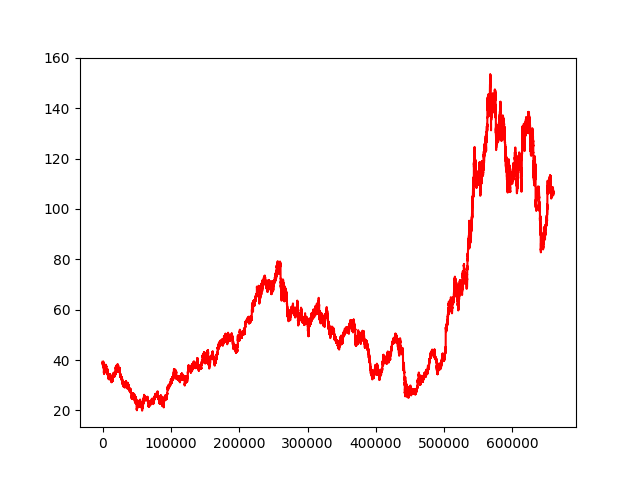}
    \end{minipage}
    \begin{minipage}{0.3\textwidth}
    \centering
    \includegraphics[width=\linewidth]{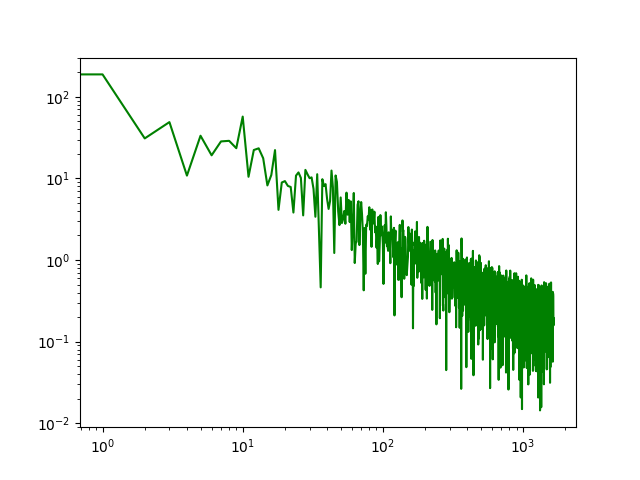}
    \end{minipage}
    \begin{minipage}{0.3\textwidth}
    \centering
    \includegraphics[width=\linewidth]{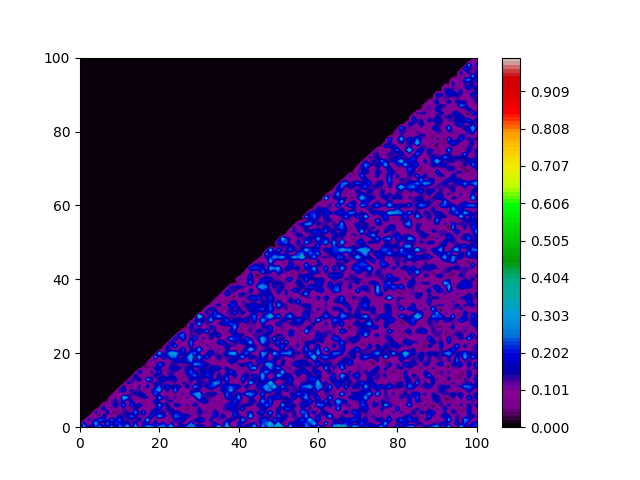}
    \end{minipage}
\caption{Raw data, turbulence spectrum and the extended integral transform of Tata Steel Index data.}
\label{Tata_Steel}
\end{figure}

\begin{figure}[h]
\centering
    \begin{minipage}{0.3\textwidth}
    \centering
    \includegraphics[width=\linewidth]{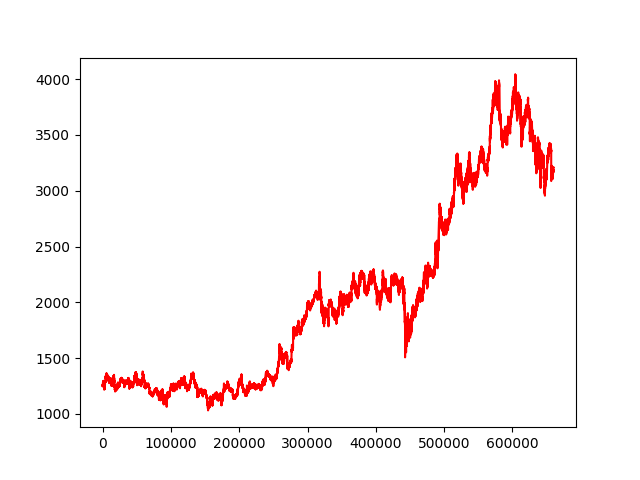}
    \end{minipage}
    \begin{minipage}{0.3\textwidth}
    \centering
    \includegraphics[width=\linewidth]{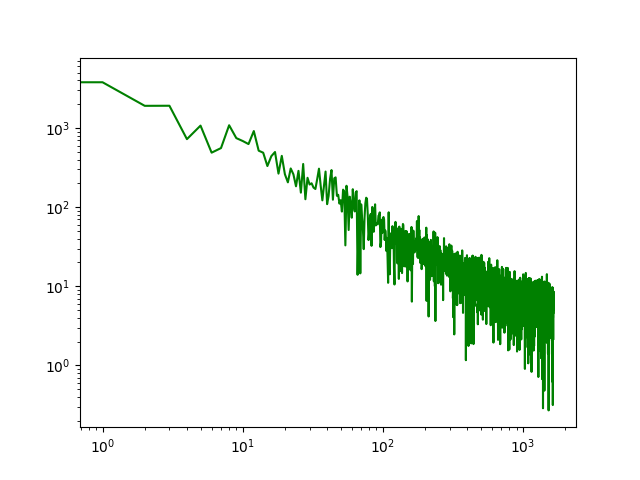}
    \end{minipage}
    \begin{minipage}{0.3\textwidth}
    \centering
    \includegraphics[width=\linewidth]{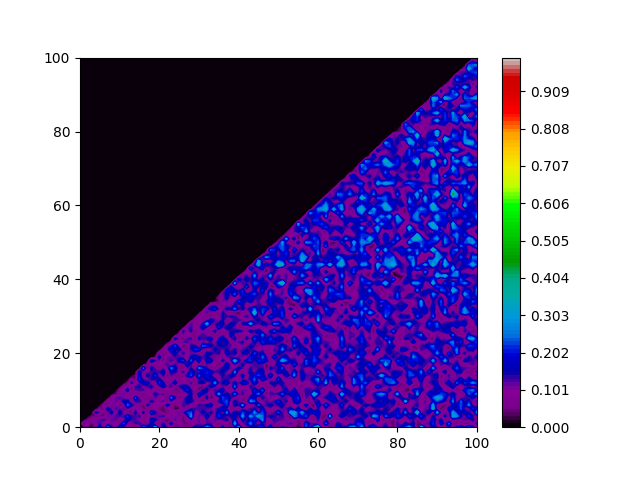}
    \end{minipage}
\caption{Raw data, turbulence spectrum and the extended integral transform of TCS Index data.} 
\label{TCS}
\end{figure}
\end{document}